\documentclass[preprint,prd,aps,nofootinbib]{revtex4}
\usepackage{graphicx}
\UseRawInputEncoding

\begin{document}


\title{The flavor-dependent $U(1)_F$ model}

\author{Jin-Lei Yang$^{1,2,3}$\footnote{jlyang@hbu.edu.cn},Hai-Bin Zhang$^{1,2,3}$\footnote{hbzhang@hbu.edu.cn},Tai-Fu Feng$^{1,2,3}$\footnote{fengtf@hbu.edu.cn}}

\affiliation{Department of Physics, Hebei University, Baoding, 071002, China$^1$\\
Key Laboratory of High-precision Computation and Application of Quantum Field Theory of Hebei Province, Baoding, 071002, China$^2$\\
Research Center for Computational Physics of Hebei Province, Baoding, 071002, China$^3$}

\begin{abstract}
A flavor-dependent model (FDM) is proposed in this work. The model extends the Standard Model by an extra $U(1)_F$ local gauge group, two scalar doublets, one scalar singlet and two right-handed neutrinos, where the additional $U(1)_F$ charges are related to the particles' flavor. The new fermion sector in the FDM can explain the flavor mixings puzzle and the mass hierarchy puzzle simultaneously, and the nonzero Majorana neutrino masses can be obtained naturally by the Type I see-saw mechanism. In addition, the $B$ meson rare decay processes $\bar B \to X_s\gamma$, $B_s^0 \to \mu^+\mu^-$, the top quark rare decay processes $t\to ch$, $t\to uh$ and the $\tau$ lepton flavor violation processes $\tau\to 3e$, $\tau\to 3\mu$, $\mu\to 3e$ predicted in the FDM are analyzed.
\end{abstract}

\maketitle

\section{Introduction}\label{sec1}

The Standard Model (SM) achieves great success in describing the interactions of fundamental particles, and most of the observations coincide well with the SM predictions. However, the Yukawa couplings in the SM are still enigmatic, because the flavor mixings of quarks described by the Cabibbo-Kobayashi-Maskawa (CKM) matrix~\cite{Cabibbo:1963yz,Kobayashi:1973fv} are not predicted from first principles in the SM, which is the so-called flavor puzzle. And the fermions of the three families have largely distinct masses
\begin{eqnarray}
&&\frac{m_t}{m_c}\approx104,\;\frac{m_c}{m_u}\approx773,\;\frac{m_b}{m_s}\approx51,\;\frac{m_s}{m_d}\approx20,\;\frac{m_\tau}{m_\mu}\approx17,\;\frac{m_\mu}{m_e}\approx207.\label{eq1}
\end{eqnarray}
It exhibits the large hierarchical structure of masses across the three families while they share common SM gauge group quantum numbers, which is the so-called mass hierarchy puzzle. Meanwhile, the nonzero neutrino masses and mixings observed at the neutrino oscillation experiments~\cite{ParticleDataGroup:2022pth} make the so-called flavor puzzle in the SM more acutely. All of these indicate that explaining the observed fermionic mass spectrum and mixings is one of the most enigmatic questions in particle physics, which may help understand the flavor nature and seek possible new physics (NP).

In literatures, there are some attempts to explain the fermionic mass hierarchies and flavor mixings. For example, the authors of Refs.~\cite{Froggatt:1978nt,Koide:1982ax,Leurer:1992wg,Ibanez:1994ig,Babu:1995hr} try to explain the fermionic flavor mixings by imposing the family symmetries, some proposals to explain the fermionic mass hierarchies can be found in Refs.~\cite{Berezhiani:1990wn,Berezhiani:1990jj,Sakharov:1994pr,Randall:1999ee,Kaplan:2001ga,Chen:2008tc,Buras:2011ph,King:2013eh,King:2014nza,King:2015aea,King:2017guk,Weinberg:2020zba,
Feruglio:2019ybq,Abbas:2018lga,Mohanta:2022seo,Mohanta:2023soi,Abbas:2022zfb}. The three-Higgs doublet model is motivated to account for the flavor structure with three generations of fermions~\cite{Weinberg:1976hu,Lavoura:2007dw,Ivanov:2012ry,GonzalezFelipe:2013xok,GonzalezFelipe:2013yhh,Keus:2013hya,Ivanov:2014doa,Buskin:2021eig,Izawa:2022viu}, in which the number of Higgs doublets is three and has a flavor symmetry. A flavor-dependent $U(1)$ extension of the SM is proposed in Ref.~\cite{VanLoi:2023utt}, the authors explained the small mixing at quark sector by introducing a NP cut-off scale parameter. Weinberg proposed a new mechanism in which only the third generation of quarks and leptons achieve the masses at the tree level, while the masses for the second and first generations are produced by one-loop and two-loop radiative corrections respectively~\cite{Weinberg:2020zba}. However, his analysis indicated that the ratios of the masses of the second and third generations fermions are independent of various masses of the third generation, i.e.
$$\frac{m_c}{m_t}=\frac{m_s}{m_b}=\frac{m_\mu}{m_\tau}\;\;{\rm or}\;\;\frac{m_c}{m_t^3}=\frac{m_s}{m_b^3}=\frac{m_\mu}{m_\tau^3},\\$$
which is not true of observed masses as shown in Eq.~(\ref{eq1}). Hence, he pointed out in his work that ``this kind of NP models are not realistic for some reasons''~\cite{Weinberg:2020zba}. In addition, Weinberg was the first to propose the Type I see-saw mechanism to give the tiny neutrino masses naturally~\cite{Weinberg:1979sa}, which provides one of the most popular mechanisms so far to give the tiny Majorana neutrino masses.

In this work, we adopt the Weinberg's idea ``only the third generation of fermions achieve the masses at the tree level'', but abandon ``the masses for the second and first generations are produced by one-loop and two-loop radiative corrections respectively''. Instead, we propose to obtain the masses for the second and first generations by the ``see-saw mechanism'' (which is proposed by Weinberg to give the tiny Majorana neutrino masses as mentioned above), i.e. the masses for the second and first generations are produced by mixing with the third generation. In this case, the ratios of the masses of the second and third generations fermions depend on the tree-level mixing parameters of fermions, and we propose a flavor-dependent model (FDM) to realize this mechanism. The model extends the SM by an extra $U(1)_F$ local gauge group which relates to the particles' flavor, two scalar doublets, one scalar singlet and two right-handed neutrinos. The new fermion sector in the FDM relates the fermionic flavor mixings to the fermionic mass hierarchies, which provides a new understanding about the mass hierarchy puzzle and the flavor mixing puzzle. In addition, the nonzero neutrino masses can be obtained naturally in the FDM by the Type I see-saw mechanism.

The paper is organized as follows: The structure of the FDM including particle content, scalar sector, fermion masses and gauge sector are collected in Sec.~\ref{sec2}. The numerical results of CKM matrix and Higgs masses predicted in the FDM are presented in Sec.~\ref{sec3}. The processes mediated by the flavor changed neutral currents (FCNCs) in the FDM are analyzed in Sec.~\ref{sec4}. The summary is made in Sec.~\ref{sec5}.

\section{The flavor-dependent model}\label{sec2}

\begin{table*}
\begin{tabular*}{\textwidth}{@{\extracolsep{\fill}}lllll@{}}
\hline
Multiplets & $SU(3)_C$ & $SU(2)_L$ & $U(1)_Y$ & $U(1)_F$\\
\hline
$l_1=(\nu_{1L},e_{1L})^T$ & 1 & 2 & $-\frac{1}{2}$ & $z$\\
$l_2=(\nu_{2L},e_{2L})^T$ & 1 & 2 & $-\frac{1}{2}$ & $-z$\\
$l_3=(\nu_{3L},e_{3L})^T$ & 1 & 2 & $-\frac{1}{2}$ & $0$\\
$\nu_{1R}$                & 1 & 1 & $0$            & $-z$ \\
$\nu_{2R}$                & 1 & 1 & $0$            & $z$ \\
$e_{1R}$                  & 1 & 1 & $-1$           & $-z$ \\
$e_{2R}$                  & 1 & 1 & $-1$           & $z$ \\
$e_{3R}$                  & 1 & 1 & $-1$           & $0$ \\
$q_1=(u_{1L},d_{1L})^T$   & 3 & 2 & $\frac{1}{6}$  & $z$\\
$q_2=(u_{2L},d_{2L})^T$   & 3 & 2 & $\frac{1}{6}$  & $-z$\\
$q_3=(u_{3L},d_{3L})^T$   & 3 & 2 & $\frac{1}{6}$  & $0$\\
$d_{1R}$                  & 3 & 1 & -$\frac{1}{3}$ & $-z$ \\
$d_{2R}$                  & 3 & 1 & -$\frac{1}{3}$ & $z$ \\
$d_{3R}$                  & 3 & 1 & -$\frac{1}{3}$ & $0$ \\
$u_{1R}$                  & 3 & 1 & $\frac{2}{3}$  & $-z$ \\
$u_{2R}$                  & 3 & 1 & $\frac{2}{3}$  & $z$ \\
$u_{3R}$                  & 3 & 1 & $\frac{2}{3}$  & $0$ \\
$\Phi_1=(\phi_1^{+},\phi_1^{0})^T$   & 1 & 2 & $\frac{1}{2}$  & $z$\\
$\Phi_2=(\phi_2^{+},\phi_2^{0})^T$   & 1 & 2 & $\frac{1}{2}$  & $-z$\\
$\Phi_3=(\phi_3^{+},\phi_3^{0})^T$   & 1 & 2 & $\frac{1}{2}$  & 0\\
$\chi$   & 1 & 1 & 0  & $2z$\\
\hline
\end{tabular*}
\caption{Matter content in the FDM, where the nonzero constant $z$ denotes the extra $U(1)_F$ charge.}
\label{tab2}
\end{table*}

The gauge group of the FDM is $SU(3)_C\otimes SU(2)_L\otimes U(1)_Y\otimes U(1)_F$, where the extra $U(1)_F$ local gauge group is related to the particles' flavor. In the FDM, the third generation of fermions obtain masses through the tree-level couplings with the SM scalar doublet, and the first two generations of fermions achieve masses through the tree-level mixings with the third generation as mentioned above. Hence, two additional scalar doublets are introduced in the FDM to realize the tree-level mixings of the first two generations and the third generation. In addition, to coincide with the observed neutrino oscillations, two right-handed neutrinos and one scalar singlet are introduced. Then the right-handed neutrinos obtain large Majorana masses after the scalar singlet achieving large vacuum expectation value (VEV), and the tiny neutrino masses and neutrino flavor mixings can be obtained by the Type I see-saw mechanism.

All fields in the FDM and the corresponding gauge symmetry charges are presented in Tab.~\ref{tab2}, where $\Phi_3$ corresponds to the SM Higgs doublet, the nonzero constant $z$ denotes the extra $U(1)_F$ charge. It can be noted in Tab.~\ref{tab2} that there are only two generations of right-handed neutrinos in the FDM, because both $U(1)_F$ and $U(1)_Y$ charges of the third generation of right-handed neutrinos $\nu_{R_3}$ are zero, which is trivial. In addition, it is obvious that the chiral anomaly cancellation can be guaranteed for the fermionic charges presented in Tab.~\ref{tab2}.

\subsection{The scalar sector of the FDM}\label{sec2-1}

The scalar potential in the FDM can be written as
\begin{eqnarray}
&&V=-M_{\Phi_1}^2 \Phi_1^\dagger\Phi_1-M_{\Phi_2}^2 \Phi_2^\dagger\Phi_2-M_{\Phi_3}^2 \Phi_3^\dagger\Phi_3-M_{\chi}^2\chi^*\chi+\lambda_{\chi} (\chi^*\chi)^2+\lambda_1 (\Phi_1^\dagger\Phi_1)^2\nonumber\\
&&\qquad+\lambda_2 (\Phi_2^\dagger\Phi_2)^2+\lambda_3 (\Phi_3^\dagger\Phi_3)^2+\lambda'_4 (\Phi_1^\dagger\Phi_1)(\Phi_2^\dagger\Phi_2)+\lambda_4'' (\Phi_1^\dagger\Phi_2)(\Phi_2^\dagger\Phi_1)\nonumber\\
&&\qquad+\lambda_5' (\Phi_1^\dagger\Phi_1)(\Phi_3^\dagger\Phi_3)+\lambda_5'' (\Phi_1^\dagger\Phi_3)(\Phi_3^\dagger\Phi_1)+\lambda_6' (\Phi_2^\dagger\Phi_2)(\Phi_3^\dagger\Phi_3)+\lambda_6'' (\Phi_2^\dagger\Phi_3)(\Phi_3^\dagger\Phi_2)\nonumber\\
&&\qquad+\lambda_7 (\Phi_1^\dagger\Phi_1)(\chi^*\chi)+\lambda_{8} (\Phi_2^\dagger\Phi_2)(\chi^*\chi)+\lambda_{9} (\Phi_3^\dagger\Phi_3)(\chi^*\chi)+[\lambda_{10} (\Phi_3^\dagger\Phi_1)(\Phi_3^\dagger\Phi_2)\nonumber\\
&&\qquad+\kappa(\Phi_1^\dagger\Phi_2)\chi+h.c.],\label{eqsca}
\end{eqnarray}
where
\begin{eqnarray}
&&\Phi_1=\left(\begin{array}{c}\phi_1^+\\ \frac{1}{\sqrt2}(i A_1+S_1+v_1)\end{array}\right),\Phi_2=\left(\begin{array}{c}\phi_2^+\\ \frac{1}{\sqrt2}(i A_2+S_2+v_2)\end{array}\right),\Phi_3=\left(\begin{array}{c}\phi_3^+\\ \frac{1}{\sqrt2}(i A_3+S_3+v_3)\end{array}\right),\nonumber\\
&&\chi=\frac{1}{\sqrt2}(i A_{\chi}+S_{\chi}+v_\chi),
\end{eqnarray}
and $v_i\;(i=1,\;2,\;3),\;v_\chi$ are the VEVs of $\Phi_i,\;\chi$ respectively.

Based on the scalar potential in Eq.~(\ref{eqsca}), the tadpole equations in the FDM can be written as\footnote{Calculating the exact vacuum stability conditions for any new physics model is difficult generally, and the obtained tadpole equations can be used to calculate the stationary points. In this case, we apply tadpole equations in the calculations, and guarantee the stability of vacuum numerically by keeping the scalar potential at the input $v_1,\;v_2,\;v_3,\;v_\chi$ are smaller than all the other stationary points.}
\begin{eqnarray}
&&M_{\Phi_1}^2=\lambda_1v_1^2+\frac{1}{2}\Big[(\lambda_4'+\lambda_4'') v_2^2+(\lambda_5'+\lambda_5'') v_3^2+\frac{v_2}{v_1}v_3^2 {\rm Re}(\lambda_{10})+\sqrt2\frac{v_2}{v_1} v_\chi {\rm Re}(\kappa)+\lambda_7v_\chi^2\Big],\nonumber\\
&&M_{\Phi_2}^2=\lambda_2v_2^2+\frac{1}{2}\Big[(\lambda_4'+\lambda_4'') v_1^2+(\lambda_6'+\lambda_6'') v_3^2+\frac{v_1}{v_2}v_3^2 {\rm Re}(\lambda_{10})+\sqrt2\frac{v_1}{v_2} v_\chi {\rm Re}(\kappa)+\lambda_8v_\chi^2\Big],\nonumber\\
&&M_{\Phi_3}^2=\lambda_3v_3^2+{\rm Re}(\lambda_{10})v_1v_2+\frac{1}{2}[(\lambda_5'+\lambda_5'') v_1^2+(\lambda_6'+\lambda_6'') v_2^2+\lambda_9 v_c^2],\nonumber\\
&&M_{\chi}^2=\lambda_\chi v_\chi^2+\frac{1}{2}\Big[\lambda_7 v_1^2+\lambda_8 v_2^2+\lambda_9 v_3^2+\sqrt2\frac{v_1v_2}{v_\chi}  {\rm Re}(\kappa)\Big].\label{eqtad}
\end{eqnarray}
On the basis $(S_1,\;S_2,\;S_3,\;S_\chi)$, the CP-even Higgs squared mass matrix in the FDM is
\begin{eqnarray}
&&M_{h}^2=\left(\begin{array}{*{20}{cccc}}
M_{h,11}^2 & M_{h,12}^2 & M_{h,13}^2 & M_{h,14}^2 \\ [6pt]
M_{h,12}^2 & M_{h,22}^2 & M_{h,23}^2 & M_{h,24}^2 \\ [6pt]
M_{h,13}^2 & M_{h,23}^2 & M_{h,33}^2 & M_{h,34}^2 \\ [6pt]
M_{h,14}^2 & M_{h,24}^2 & M_{h,34}^2 & M_{h,44}^2 \\ [6pt]
\end{array}\right),
\end{eqnarray}
where
\begin{eqnarray}
&&M_{h,11}^2=2\lambda_1v_1^2-\frac{v_2}{2v_1}\Big[v_3^2 {\rm Re}(\lambda_{10})+\sqrt2 v_\chi {\rm Re}(\kappa)\Big],\nonumber\\
&&M_{h,22}^2=2\lambda_2v_2^2-\frac{v_1}{2v_2}\Big[v_3^2 {\rm Re}(\lambda_{10})+\sqrt2 v_\chi {\rm Re}(\kappa)\Big],\nonumber\\
&&M_{h,33}^2=2\lambda_3v_3^2,\;\;M_{h,44}^2=2\lambda_\chi v_\chi^2-\frac{\sqrt2v_1v_2}{2v_\chi}  {\rm Re}(\kappa),\nonumber\\
&&M_{h,12}^2=(\lambda_4'+\lambda_4'') v_1v_2+\frac{1}{2}{\rm Re}(\lambda_{10})v_3^2+\frac{\sqrt2}{2}{\rm Re}(\kappa)v_\chi,\nonumber\\
&&M_{h,13}^2=(\lambda_5'+\lambda_5'')v_1v_3+{\rm Re}(\lambda_{10})v_2v_3,\;\;M_{h,14}^2=\lambda_7v_1v_\chi+\frac{\sqrt2}{2}{\rm Re}(\kappa)v_2,\nonumber\\
&&M_{h,23}^2=(\lambda_6'+\lambda_6'')v_2v_3+{\rm Re}(\lambda_{10})v_1v_3,\;\;M_{h,24}^2=\lambda_8v_2v_\chi+\frac{\sqrt2}{2}{\rm Re}(\kappa)v_1,\nonumber\\
&&M_{h,34}^2=\lambda_9v_3v_\chi.\label{eqmh}
\end{eqnarray}
The tadpole equations in Eq.~(\ref{eqtad}) are used to obtain the matrix elements above.

Then, on the basis $(A_1,\;A_2,\;A_3,\;A_\chi)$, the squared mass matrix of CP-odd Higgs in the FDM can be written as
\begin{eqnarray}
&&M_{A}^2=\left(\begin{array}{*{20}{cccc}}
M_{A,11}^2 & M_{A,12}^2 & M_{A,13}^2 & M_{A,14}^2 \\ [6pt]
M_{A,12}^2 & M_{A,22}^2 & M_{A,23}^2 & M_{A,24}^2 \\ [6pt]
M_{A,13}^2 & M_{A,23}^2 & M_{A,33}^2 & M_{A,34}^2 \\ [6pt]
M_{A,14}^2 & M_{A,24}^2 & M_{A,34}^2 & M_{A,44}^2 \\ [6pt]
\end{array}\right),
\end{eqnarray}
where
\begin{eqnarray}
&&M_{A,11}^2=-\frac{v_2}{2v_1}[{\rm Re}(\lambda_{10})v_3^2+\sqrt2 v_\chi {\rm Re} (\kappa)],\;\;M_{A,33}^2=-2{\rm Re}(\lambda_{10})v_1v_2,\nonumber\\
&&M_{A,22}^2=-\frac{v_1}{2v_2}[{\rm Re}(\lambda_{10})v_3^2+\sqrt2 v_\chi {\rm Re} (\kappa)],\;\;M_{A,44}^2=-\frac{\sqrt2 v_1v_2}{2v_\chi}{\rm Re} (\kappa),\nonumber\\
&&M_{A,12}^2=\frac{\sqrt 2}{2} v_\chi {\rm Re} (\kappa)-\frac{1}{2}{\rm Re}(\lambda_{10})v_3^2,\;\;M_{A,13}^2={\rm Re}(\lambda_{10})v_2v_3,\nonumber\\
&&M_{A,14}^2=\frac{\sqrt 2}{2} v_2 {\rm Re} (\kappa),\;\;M_{A,23}^2={\rm Re}(\lambda_{10})v_1v_3,\;\;M_{A,24}^2=-\frac{\sqrt 2}{2} v_1 {\rm Re} (\kappa),\nonumber\\
&&M_{A,34}^2=0.\label{eqmA}
\end{eqnarray}
On the basis $(\phi_1^+,\;\phi_2^+,\;\phi_3^+)$ and $(\phi_1^-,\;\phi_2^-,\;\phi_3^-)^T$, the squared mass matrix of singly charged Higgs in the FDM can be written as
\begin{eqnarray}
&&M_{H^\pm}^2=\left(\begin{array}{*{20}{ccc}}
M_{H^\pm,11}^2 & M_{H^\pm,12}^2 & M_{H^\pm,13}^2 \\ [6pt]
(M_{H^\pm,12}^2)^* & M_{H^\pm,22}^2 & M_{H^\pm,23}^2 \\ [6pt]
(M_{H^\pm,13}^2)^* & (M_{H^\pm,23}^2)^* & M_{H^\pm,33}^2 \\ [6pt]
\end{array}\right),
\end{eqnarray}
where
\begin{eqnarray}
&&M_{H^\pm,11}^2=-\frac{v_2}{2v_1}[{\rm Re}(\lambda_{10})v_3^2+\sqrt2 v_\chi {\rm Re} (\kappa)]-\frac{1}{2}(\lambda_4'' v_2^2+\lambda_5'' v_3^2),\nonumber\\
&&M_{H^\pm,22}^2=-\frac{v_1}{2v_2}[{\rm Re}(\lambda_{10})v_3^2+\sqrt2 v_\chi {\rm Re} (\kappa)]-\frac{1}{2}(\lambda_4'' v_1^2+\lambda_6'' v_3^2),\nonumber\\
&&M_{H^\pm,33}^2=-{\rm Re}(\lambda_{10})v_1v_2-\frac{1}{2}(\lambda_5'' v_1^2+\lambda_6'' v_2^2),\nonumber\\
&&M_{H^\pm,12}^2=\frac{\sqrt 2}{2}v_\chi\kappa+\frac{1}{2}\lambda_4'' v_1 v_2,\;\;M_{H^\pm,13}^2=\frac{1}{2}v_3(\lambda_5'' v_1+\lambda_{10}^*v_2),\nonumber\\
&&M_{H^\pm,23}^2=\frac{1}{2}v_3(\lambda_6'' v_2+\lambda_{10}^*v_1).\label{eqmCH}
\end{eqnarray}
It is easy to verify that there are two neutral Goldstones and one singly charged Goldstone in the FDM.

\subsection{The fermion masses in the FDM}\label{sec2-2}

Based on the matter content listed in Tab.~\ref{tab2}, the Yukawa couplings in the FDM can be written as
\begin{eqnarray}
&&\mathcal{L}_Y=Y_u^{33}\bar q_3 \tilde \Phi_3 u_{R_3}+Y_d^{33}\bar q_3 \Phi_3 d_{R_3}+Y_u^{32}\bar q_3 \tilde{\Phi}_1 u_{R_2}+Y_u^{23}\bar q_2 \tilde \Phi_1 u_{R_3}+Y_d^{32}\bar q_3 \Phi_2 d_{R_2}\nonumber\\
&&\qquad\; +Y_d^{23}\bar q_2 \Phi_2 d_{R_3}+Y_u^{21}\bar q_2 \tilde{\Phi}_3 u_{R_1}+Y_u^{12}\bar q_1 \tilde \Phi_3 u_{R_2}+Y_d^{21}\bar q_2 \Phi_3 d_{R_1}+ Y_d^{12}\bar q_1 \Phi_3 d_{R_2}\nonumber\\
&&\qquad\; +Y_u^{31}\bar q_3 \tilde{\Phi}_2 u_{R_1}+Y_u^{13}\bar q_1 \tilde \Phi_2 u_{R_3}+Y_d^{31}\bar q_3 \Phi_1 d_{R_1}+Y_d^{13}\bar q_1 \Phi_1 d_{R_3}\nonumber\\
&&\qquad\; +Y_e^{33}\bar l_3 \Phi_3 e_{R_3}+Y_e^{32}\bar l_3 \Phi_2 e_{R_2}+Y_e^{23}\bar l_2 \Phi_2 e_{R_3}+Y_e^{21}\bar l_2 \Phi_3 e_{R_1}+ Y_e^{12}\bar l_1 \Phi_3 e_{R_2}\nonumber\\
&&\qquad\; +Y_e^{31}\bar l_3 \Phi_1 e_{R_1}+Y_e^{13}\bar l_1 \Phi_1 e_{R_3}+Y_R^{11}\bar\nu^c_{R_1}\nu_{R_1}\chi+Y_R^{22}\bar\nu^c_{R_2}\nu_{R_2} \chi^*+Y_D^{21}\bar l_2 \tilde \Phi_3 \nu_{R_1}\nonumber\\
&&\qquad\; +Y_D^{12}\bar l_1 \tilde \Phi_3 \nu_{R_2}+Y_D^{31}\bar l_3 \tilde \Phi_2 \nu_{R_1}+Y_D^{32}\bar l_3 \tilde \Phi_1 \nu_{R_2}+h.c.,\label{eq9}
\end{eqnarray}
Then the mass matrices of quarks and leptons can be written as
\begin{eqnarray}
&&m_q=\left(\begin{array}{ccc} 0 & m_{q,12} & m_{q,13}\\
m_{q,12}^* & 0 & m_{q,23}\\
m_{q,13}^* & m_{q,23}^* & m_{q,33}\end{array}\right),m_e=\left(\begin{array}{ccc} 0 & m_{e,12} & m_{e,13}\\
m_{e,12}^* & 0 & m_{e,23}\\
m_{e,13}^* & m_{e,23}^* & m_{e,33}\end{array}\right),m_\nu=\left(\begin{array}{cc} 0 & M_D^T\\
M_D & M_R\end{array}\right),\label{eq2}
\end{eqnarray}
where $q=u,d$, the parameters $m_{q,33}$ and $m_{e,33}$ are real, $M_D$ is $2\times3$ Dirac mass matrix and $M_R$ is $2\times2$ Majorana mass matrix (the nonzero neutrino masses are obtained by the Type I see-saw mechanism). The elements of the matrices in Eq.~(\ref{eq2}) are
\begin{eqnarray}
&&m_{u,11}=m_{u,22}=0,\;m_{u,33}=\frac{1}{\sqrt2}Y_u^{33}v_3,\;m_{u,12}=\frac{1}{\sqrt2}Y_u^{12}v_3,\;m_{u,13}=\frac{1}{\sqrt2}Y_u^{13}v_1,\nonumber\\
&&m_{u,23}=\frac{1}{\sqrt2}Y_u^{23}v_2,\label{eqmu}\\
&&m_{d,11}=m_{d,22}=0,\;m_{d,33}=\frac{1}{\sqrt2}Y_d^{33}v_3,\;m_{d,12}=\frac{1}{\sqrt2}Y_d^{12}v_3,\;m_{d,13}=\frac{1}{\sqrt2}Y_d^{13}v_1,\nonumber\\
&&m_{d,23}=\frac{1}{\sqrt2}Y_d^{23}v_2,\label{eqmd}\\
&&m_{e,11}=m_{e,22}=0,\;m_{e,33}=\frac{1}{\sqrt2}Y_e^{33}v_3,\;m_{e,12}=\frac{1}{\sqrt2}Y_e^{12}v_3,\;m_{e,13}=\frac{1}{\sqrt2}Y_e^{13}v_1,\nonumber\\
&&m_{e,23}=\frac{1}{\sqrt2}Y_e^{23}v_2,\label{eqme}\\
&&M_{D,11}=M_{D,22}=0,\;\;M_{D,12}=\frac{1}{\sqrt2}Y_D^{12}v_3,\;M_{D,31}=\frac{1}{\sqrt2}Y_D^{31}v_1,\nonumber\\
&&M_{D,32}=\frac{1}{\sqrt2}Y_D^{32}v_2,\;M_{R,12}=M_{R,21}=0,\;M_{R,11}=\frac{1}{\sqrt2}Y_R^{11}v_\chi,\;M_{R,22}=\frac{1}{\sqrt2}Y_R^{22}v_\chi.
\end{eqnarray}

\subsection{The gauge sector of the FDM}\label{sec2-3}

Due to the introducing of an extra $U(1)_F$ local gauge group in the FDM, the covariant derivative corresponding to $SU(2)_L\otimes U(1)_Y\otimes U(1)_F$ is defined as
\begin{eqnarray}
&&D_\mu=\partial_\mu+i g_2 T_j A_{j\mu}+i g_1 Y B_\mu+i g_{_F} F B'_\mu+i g_{_{YF}} Y B'_\mu,\;(j=1,\;2,\;3),\label{eqCD}
\end{eqnarray}
where $(g_2,\;g_1,\; g_{_F})$, $(T_j,\;Y,\;F)$, $(A_{j\mu},\;B_\mu,\; B'_\mu)$ denote the gauge coupling constants, generators and gauge bosons of groups $(SU(2)_L,\;U(1)_Y,\;U(1)_F)$ respectively, $g_{_{YF}}$ is the gauge coupling constant arises from the gauge kinetic mixing effect which presents in the models with two Abelian groups. Then the $W$ boson mass can be written as
\begin{eqnarray}
&&M_W=\frac{1}{2}g_2 (v_1^2+v_2^2+v_3^2)^{1/2},
\end{eqnarray}
where $(v_1^2+v_2^2+v_3^2)^{1/2}=v\approx246\;{\rm GeV}$ and we have $v_1,\;v_2 < v_3$ in the FDM. The $\gamma$, $Z$ and $Z'$ boson masses in the FDM can be written as
\begin{eqnarray}
&&M_\gamma=0,\;M_Z\approx\frac{1}{2}(g_1^2+g_2^2)^{1/2} v,\;M_{Z'}\approx 2|zg_{_F}| v_\chi,\label{eq19}
\end{eqnarray}
and
\begin{eqnarray}
&&\gamma=c_W B+s_W A_3,\;Z=-s_W B+c_W A_3+s'_W B',\;Z'=s_W'(s_WB-c_W A_3)+c'_W B',
\end{eqnarray}
where $\gamma,\;Z,\;Z'$ are the mass eigenstates, $c_W\equiv \cos \theta_W,\;s_W\equiv \sin \theta_W$ with $\theta_W$ denoting the Weinberg angle, $s_W'\equiv \sin \theta'_W$, $c_W'\equiv \cos \theta'_W$ with $\theta_W'$ representing the $Z-Z'$ mixing effect.

\section{CKM matrix and Higgs masses in the FDM}\label{sec3}

The quark sector in the FDM are redefined and the additional two scalar doublets, one scalar singlet modify the scalar potential of the model, hence we focus on the quark sector and scalar sector of the model in this section.

\subsection{CKM matrix}\label{sec3-1}

The analysis in our previous work~\cite{Yang:2024kfs} shows that the quark mass matrices obtained in the FDM can fit the measured CKM matrix well. In this work, we perform a $\chi^2$ test to explore the best fit describing the quark masses and CKM matrix in the model. Generally, the $\chi^2$ function can be constructed as
\begin{eqnarray}
&&\chi^2=\sum_1 \Big(\frac{O_i^{\rm th}-O_i^{\rm exp}}{\sigma_i^{\rm exp}}\Big)^2,
\end{eqnarray}
where $O_i^{\rm th}$ denotes the $i-$th observable computed theoretically, $O_i^{\rm exp}$ is the corresponding experimental value and $\sigma_i^{\rm exp}$ is the uncertainty in $O_i^{\rm exp}$. Taking into account $10$ observables including $6$ quark masses, $3$ mixing angles and a phase in the CKM matrix, we scan the parameter space
\begin{eqnarray}
&&|m_{u,13}|=(0.0\sim1.0)\;{\rm GeV},\;|m_{d,13}|=(0.0\sim0.1)\;{\rm GeV},\;\theta_{q,ij}=(-\pi,\pi)\label{eqSq}
\end{eqnarray}
with $m_{q,ij}=|m_{q,ij}|e^{i\theta_{q,ij}}$, $q=u,d$, $ij=12,13,23$. Then we obtain the best fit solution corresponding to $\chi^2=0.0072$, the results of this fit are listed in Tab.~\ref{tab3} where various $O_i^{\rm exp}$ we use are listed in the third column and $\sigma_i^{\rm exp}$ are take from PDG~\cite{ParticleDataGroup:2022pth}.
\begin{table*}
\begin{tabular*}{\textwidth}{@{\extracolsep{\fill}}llll@{}}
\hline
Observables & $O_i^{\rm th}$ & $O_i^{\rm exp}$ & Deviations in $\%$\\
\hline
$m_u$[MeV] & 2.15   & 2.16     & 0.47\\
$m_c$[GeV] & 1.67   & 1.67     & 0\\
$m_t$[GeV] & 172.5  & 172.5    & 0\\
$m_d$[MeV] & 4.67   & 4.67     & 0 \\
$m_s$[MeV] & 93.4   & 93.4     & 0 \\
$m_b$[GeV] & 4.78   & 4.78     & 0 \\
$|v_{us}|$ & 0.2253 & 0.2253   & 0 \\
$|v_{ub}|$ & 0.3616 & 0.003616 & 0 \\
$|v_{cb}|$ & 0.4149 & 0.04149  & 0\\
\hline
\end{tabular*}
\caption{The results obtained for the best fit corresponding to $\chi^2=\mathbf{0.0072}$.}
\label{tab3}
\end{table*}
The parameters for the best fit listed in Tab.~\ref{tab3} are
\begin{eqnarray}
&&|m_{u,13}|=0.3152\;{\rm GeV},\;\theta_{u,12}=-0.8248\pi,\;\theta_{u,13}=0.1231\pi,\;\theta_{u,23}=-0.1129\pi,\nonumber\\
&&|m_{d,13}|=7.235\;{\rm MeV},\;\theta_{d,12}=0.6705\pi,\;\theta_{d,13}=-0.3622\pi,\;\theta_{d,23}=-0.06389\pi,\label{eq22}
\end{eqnarray}
and $|m_{q,12}|,\;|m_{q,23}|,\;m_{q,33}\;(q=u,\;d)$ can be obtained by\footnote{Eq.~(\ref{eq3}) is obtained with the approximation $|m_{q,12}|,\;|m_{q,13}|,\;|m_{q,23}|\ll m_{q,33}$, and the terms of $\mathcal{O}(m_{q,ij}/m_{q,33})^3$ are neglected. For down type quark, higher order corrections to Eq.~(\ref{eq3}) are also important, we calculated them numerically and the corrections to $m_{q,33}$, $|m_{q,23}|$, $|m_{q,12}|$ are $9.32$ MeV, $-27.03$ MeV, $0.196$ MeV respectively. }
\begin{eqnarray}
&&m_{q,33}=m_{q_3}-m_{q_1}-m_{q_2},\nonumber\\
&&|m_{q,23}|=[(m_{q_1}+m_{q_2})m_{q,33}-|m_{q,13}|^2]^{1/2},\nonumber\\
&&|m_{q,12}|=\frac{|m_{q,13}||m_{q,23}|}{|m_{q,33}|} \cos(\theta_{q,12}+\theta_{q,23}-\theta_{q,13})\nonumber\\
&&\qquad\qquad+\Big\{[\frac{|m_{q,13}||m_{q,23}|}{|m_{q,33}|^2}\cos(\theta_{q,12}+\theta_{q,23}-\theta_{q,13})]^2+\frac{m_{q_1}m_{q_2}}{|m_{q,33}|^2}\Big\}^{1/2}|m_{q,33}|,\label{eq3}
\end{eqnarray}
with $m_{q_k}$ being the $k-$generation quark $q$ mass.

It is obvious that there are additional CP phases in the obtained CKM matrix by taking the parameters in Eq.~(\ref{eq22}) as inputs, but we do not list the observed CP phase in the CKM matrix in Tab.~\ref{tab3}. Because the CKM matrix is defined as
\begin{eqnarray}
&&V_{\rm CKM}=(U_L^{u}U_L^{d\dagger})^*,
\end{eqnarray}
where $U_L^{u}$, $U_L^{d}$ are the unitary matrices which diagonalize the quark matrices in Eq.~(\ref{eq2})
\begin{eqnarray}
&&m_u^{\rm diag}=U_L^{u*}m_uU_R^{u\dagger},\;m_d^{\rm diag}=U_L^{d*}m_dU_R^{d\dagger}.\label{eq23}
\end{eqnarray}
Eq.~(\ref{eq23}) is invariant under
\begin{eqnarray}
&&U_L^{u}\to {\rm diag}(e^{-i \theta_u},e^{-i \theta_c},e^{-i \theta_t})\cdot U_L^{u},\;U_R^{u}\to {\rm diag}(e^{i \theta_u},e^{i \theta_c},e^{i \theta_t})\cdot U_R^{u},\nonumber\\
&&U_L^{d}\to {\rm diag}(e^{i \theta_d},e^{i \theta_s},e^{i \theta_b})\cdot U_L^{d},\;U_R^{d}\to {\rm diag}(e^{-i \theta_d},e^{-i \theta_s},e^{-i \theta_b})\cdot U_R^{d},
\end{eqnarray}
so there are six free parameters $\theta_u,\;\theta_c,\;\theta_t,\;\theta_d,\;\theta_s,\;\theta_b$ which can absorb the extra CP phases in the obtained CKM matrix, these parameters are not observable. As a result, the observed CP phase in the CKM matrix can be obtained by defining appropriate $\theta_u,\;\theta_c,\;\theta_t,\;\theta_d,\;\theta_s,\;\theta_b$. This fact is always valid because there are also six possible CP violation parameters at the quark sector in the model.

\subsection{Perturbative unitary bounds}\label{sec3-2}

The scalar sector of the FDM is extended by two scalar doublets and one scalar singlet, additional new couplings are also introduced correspondingly. Therefore, the tree-level perturbative unitary should be applied to the scalar elastic scattering processes in this model~\cite{Lee:1977eg}. The zero partial wave amplitude
\begin{eqnarray}
&&a_0=\frac{1}{32\pi}\sqrt{\frac{4p_f^{\rm CM}p_i^{\rm CM}}{s}}\int_{-1}^{+1}T_{2\to2}d\cos\theta
\end{eqnarray}
must satisfy the condition $|{\rm Re}(a_0)|\leq\frac{1}{2}$, where $s$ is the centre of mass (CM) energy, $T_{2\to2}$ denotes the matrix element for $2\to2$ processes, $\theta$ is the incident angle between two incoming particles, $p_i^{\rm CM}$ and $p_f^{\rm CM}$ are the initial and final momenta in the CM system respectively.

The possible two particle states of $2\to 2$ scattering processes at the scalar sector in the FDM are $S_1S_1$, $S_2S_2$, $S_3S_3$, $S_\chi S_\chi$, $S_1S_2$, $S_1S_3$, $S_1S_\chi$, $S_2S_3$, $S_2S_\chi$, $S_3S_\chi$. Considering the limit $s\gg M_{S_1}^2,\;M_{S_2}^2,\;M_{S_3}^2,\;M_{S_\chi}^2$, we have
\begin{eqnarray}
&&a_0^{S_1S_1\to S_1S_1}\approx \frac{1}{16\pi}(6\lambda_1),\;a_0^{S_2S_2\to S_2S_2}\approx \frac{1}{16\pi}(6\lambda_2),\nonumber\\
&&a_0^{S_3S_3\to S_3S_3}\approx \frac{1}{16\pi}(6\lambda_3),\;a_0^{S_\chi S_\chi\to S_\chi S_\chi}\approx \frac{1}{16\pi}(6\lambda_\chi),\nonumber\\
&&a_0^{S_1S_1\to S_2S_2}\approx \frac{1}{16\pi}(\lambda_4'+\lambda_4''),\;a_0^{S_1S_2\to S_1S_2}\approx \frac{1}{16\pi}(\lambda_4'+\lambda_4''),\nonumber\\
&&a_0^{S_1S_1\to S_3S_3}\approx \frac{1}{16\pi}(\lambda_5'+\lambda_5''),\;a_0^{S_1S_3\to S_1S_3}\approx \frac{1}{16\pi}(\lambda_5'+\lambda_5''),\nonumber\\
&&a_0^{S_2S_2\to S_3S_3}\approx \frac{1}{16\pi}(\lambda_6'+\lambda_6''),\;a_0^{S_2S_3\to S_2S_3}\approx \frac{1}{16\pi}(\lambda_6'+\lambda_6''),\nonumber\\
&&a_0^{S_1S_1\to S_\chi S_\chi}\approx \frac{1}{16\pi}(\lambda_7),\;a_0^{S_1S_\chi\to S_1S_\chi}\approx \frac{1}{16\pi}(\lambda_7),\nonumber\\
&&a_0^{S_2S_2\to S_\chi S_\chi}\approx \frac{1}{16\pi}(\lambda_8),\;a_0^{S_2S_\chi\to S_2S_\chi}\approx \frac{1}{16\pi}(\lambda_8),\nonumber\\
&&a_0^{S_3S_3\to S_\chi S_\chi}\approx \frac{1}{16\pi}(\lambda_9),\;a_0^{S_3S_\chi\to S_3S_\chi}\approx \frac{1}{16\pi}(\lambda_9),\nonumber\\
&&a_0^{S_3S_3\to S_1 S_2}\approx \frac{1}{16\pi}[2{\rm Re}(\lambda_{10})],\;a_0^{S_1S_3\to S_2 S_3}\approx \frac{1}{16\pi}[2{\rm Re}(\lambda_{10})].
\end{eqnarray}
Requiring $|{\rm Re}(a_0)|\leq\frac{1}{2}$ for each individual process, we have
\begin{eqnarray}
&&|\lambda_1|,\;|\lambda_2|,\;|\lambda_3|,\;|\lambda_\chi|\leq\frac{4\pi}{3},\;|{\rm Re}(\lambda_{10})|\leq4\pi,\nonumber\\
&&|\lambda_4'+\lambda_4''|,\;|\lambda_5'+\lambda_5''|,\;|\lambda_6'+\lambda_6''|,\;|\lambda_7|,\;|\lambda_8|,\;|\lambda_9|\leq 8\pi.\label{eqPU}
\end{eqnarray}

\subsection{Higgs masses in the FDM}\label{sec3-3}

For the free parameters in the scalar sector of the FDM, we take $v_1=v_2$, $\lambda_4'=\lambda_4''=\lambda_4/2$, $\lambda_5'=\lambda_5''=\lambda_5/2$, $\lambda_6'=\lambda_6''=\lambda_6/2$ and all parameters to be real for simplicity. In addition, $v_\chi$ is limited by the new $Z'$ boson mass as shown in Eq.~(\ref{eq19}), hence we take $v_\chi\geq5\;{\rm TeV}$ in the following analysis. Considering the perturbative unitary bounds presented in Eq.~(\ref{eqPU}), we scan the following parameter space
\begin{eqnarray}
&&v_1=(0,\;40)\;{\rm GeV},\;\lambda_i=(0,\;4)\;\;{\rm with}\;\;(i=1,...,9,\chi),\;\lambda_{10}=(-4,\;0),\nonumber\\
&&v_\chi=(5,\;40)\;{\rm TeV},\;\kappa=(-3,\;-0.1)\;{\rm TeV},\label{eq25}
\end{eqnarray}
to explore the Higgs mass spectrum in the FDM. In the scanning, we keep the next-to-lightest CP-even Higgs mass $M_{H_2}$ in the range $124\;{\rm GeV}<M_{H_2}<126\;{\rm GeV}$ and the scalar potential at the input $v_1,\;v_2,\;v_3,\;v_\chi$ are smaller than all the other stationary points to guarantee the stability of vacuum.

\subsubsection{CP-even Higgs masses}\label{sec3-3-1}

\begin{figure}
\setlength{\unitlength}{1mm}
\centering
\includegraphics[width=2.1in]{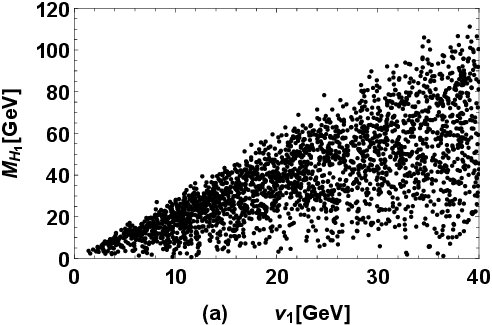}
\vspace{0.1cm}
\includegraphics[width=2.1in]{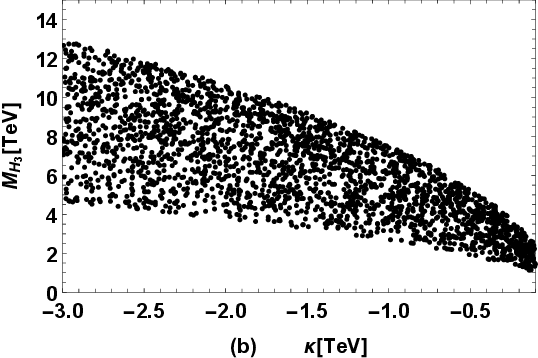}
\vspace{0.1cm}
\includegraphics[width=2.1in]{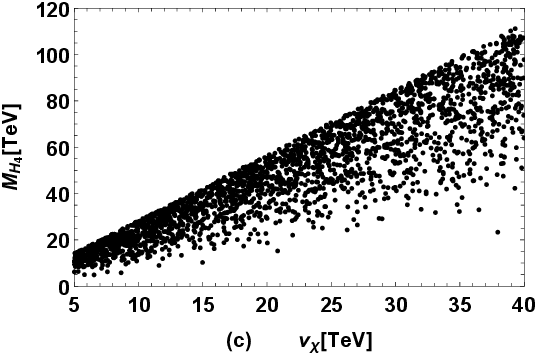}
\vspace{-0.2cm}
\caption[]{The results of CP-even Higgs masses $M_{H_1}$ (a), $M_{H_3}$ (b), $M_{H_4}$ (c) versus $v_1$, $\kappa$, $v_{\chi}$ are plotted respectively by scanning the parameter space in Eq.~(\ref{eq25}).}
\label{MHmass}
\end{figure}
The results of CP-even Higgs masses $M_{H_1}$ versus $v_1$, $M_{H_3}$ versus $\kappa$, $M_{H_4}$ versus $v_{\chi}$ are plotted in Fig.~\ref{MHmass} (a), Fig.~\ref{MHmass} (b), Fig.~\ref{MHmass} (c) respectively. We do not present the results of the next-to-lightest CP-even Higgs mass $M_{H_2}$ in Fig.~\ref{MHmass} because $M_{H_2}$ is limited in the range $124\;{\rm GeV}<M_{H_2}<126\;{\rm GeV}$ in the plotting. Fig.~\ref{MHmass} (a) shows that the lightest Higgs mass $M_{H_1}$ in the FDM mainly depends on the chosen value of $v_1$, and $M_{H_1}$ can reach $95\;{\rm GeV}$ for $v_1\gtrsim30\;{\rm GeV}$ which will be explored in detail in our next work. As shown in Fig.~\ref{MHmass} (b) and (c), $M_{H_3}$, $M_{H_4}$ are dominated by $\kappa$, $v_{\chi}$, where $M_{H_4}$ is about $110\;{\rm TeV}$ for large $v_{\chi}$ while $M_{H_3}$ is about $13\;{\rm TeV}$ for large $|\kappa|$.

\begin{figure}
\setlength{\unitlength}{1mm}
\centering
\includegraphics[width=2.7in]{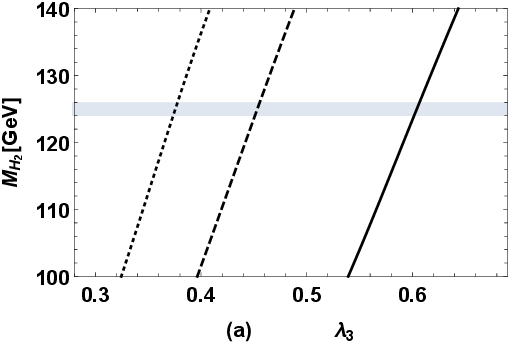}
\vspace{0.1cm}
\includegraphics[width=2.7in]{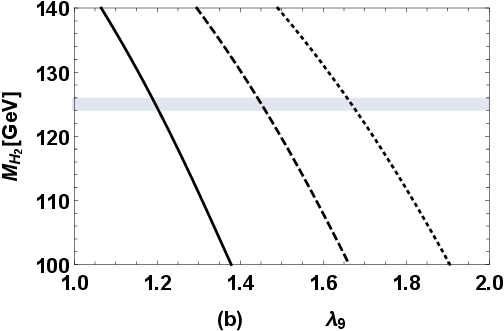}
\vspace{-0.2cm}
\caption[]{The results of $M_{H_2}$ versus $\lambda_3$ (a), $\lambda_9$ (b), where the solid, dashed, dotted curves denotes the results for $\lambda_\chi=2,\;3,\;4$ respectively, and the gray areas denote the range $124\;{\rm GeV}<M_{H_2}<126\;{\rm GeV}$.}
\label{MH2}
\end{figure}
As mentioned above, $\Phi_3$ corresponds to the SM Higgs doublet, hence $\lambda_3$ may affect the $125\;{\rm GeV}$ Higgs boson mass $M_{H_2}$ significantly. In addition, Eq.~(\ref{eqmh}) shows that there are large mixing effects between $S_3$ and $S_\chi$, it indicates $\lambda_9$, $\lambda_\chi$ can also affect $M_{H_2}$. And to verify numerically that $M_{H_2}$ is mainly affected by $\lambda_3$, $\lambda_9$, $\lambda_\chi$, we take $v_1=20\;{\rm GeV},\;\lambda_1=\lambda_2=\lambda_4=\lambda_5=\lambda_6=\lambda_7=\lambda_8=2,\;\lambda_{10}=-2,\;v_\chi=10\;{\rm TeV},\;\kappa=-1\;{\rm TeV}$ to explore the effects of $\lambda_3$, $\lambda_9$, $\lambda_\chi$ on $M_{H_2}$. Then $M_{H_2}$ versus $\lambda_3$, $\lambda_9$ are plotted in Fig.~\ref{MH2} (a) and Fig.~\ref{MH2} (b) respectively, where the solid, dashed, dotted curves denotes the results for $\lambda_\chi=2,\;3,\;4$ respectively, and the gray areas denote the range $124\;{\rm GeV}<M_{H_2}<126\;{\rm GeV}$. The picture shows $M_{H_2}$ increases with increasing $\lambda_3$, $\lambda_\chi$ and decreases with increasing $\lambda_9$, and $\lambda_3$, $\lambda_9$, $\lambda_\chi$ affect the theoretical predictions on $M_{H_2}$ significantly.

\subsubsection{CP-odd Higgs masses}\label{sec3-3-2}

\begin{figure}
\setlength{\unitlength}{1mm}
\centering
\includegraphics[width=2.1in]{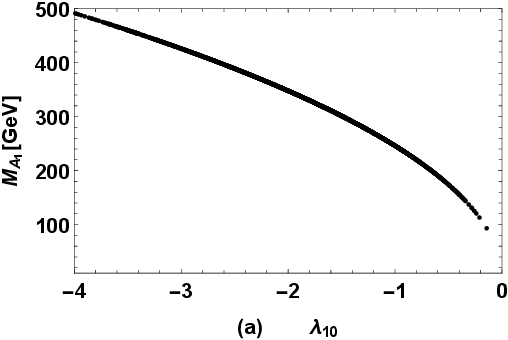}
\vspace{0.1cm}
\includegraphics[width=2.1in]{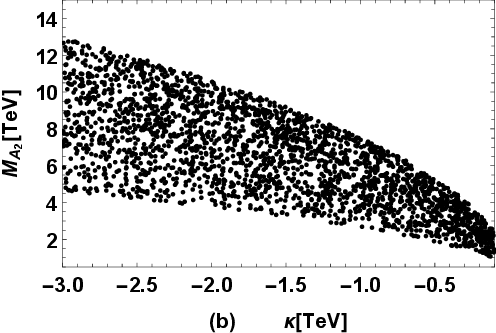}
\vspace{0.1cm}
\includegraphics[width=2.1in]{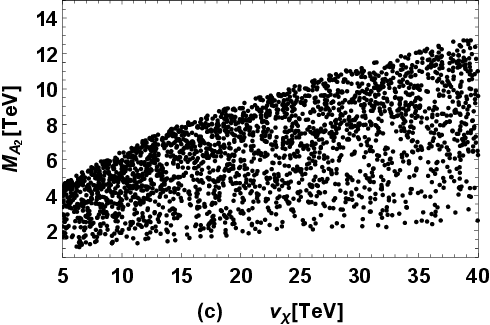}
\vspace{-0.2cm}
\caption[]{The results of CP-even Higgs masses $M_{A_1}$ (a), $M_{A_2}$ (b), $M_{A_2}$ (c) versus $\lambda_{10}$, $\kappa$, $v_\chi$ are plotted respectively by scanning the parameter space in Eq.~(\ref{eq25}).}
\label{MAmass}
\end{figure}
There are two physical CP-odd Higgs in the FDM, and the results of CP-odd Higgs masses $M_{A_1}$ versus $\lambda_{10}$, $M_{A_2}$ versus $\kappa$, $M_{A_2}$ versus $v_\chi$ are plotted in Fig.~\ref{MAmass} (a), Fig.~\ref{MAmass} (b), Fig.~\ref{MAmass} (c) respectively. It is obvious in Fig.~\ref{MAmass} (a) that $M_{A_1}$ is dominated by the value of $\lambda_{10}$ completely, and $M_{A_1}$ increases with increasing $|\lambda_{10}|$. In addition, $M_{A_2}$ is dominated by $\kappa$, $v_{\chi}$ as shown in Fig.~\ref{MAmass} (b) and Fig.~\ref{MAmass} (c), where $M_{A_2}$ can be large when $|\kappa|$ and $v_{\chi}$ are large.

\subsubsection{Charged Higgs masses}\label{sec3-3-3}

\begin{figure}
\setlength{\unitlength}{1mm}
\centering
\includegraphics[width=2.1in]{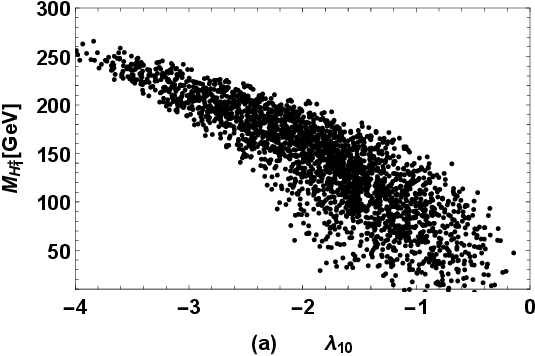}
\vspace{0.1cm}
\includegraphics[width=2.1in]{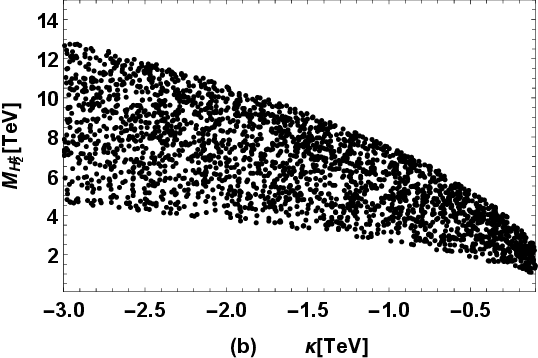}
\vspace{0.1cm}
\includegraphics[width=2.1in]{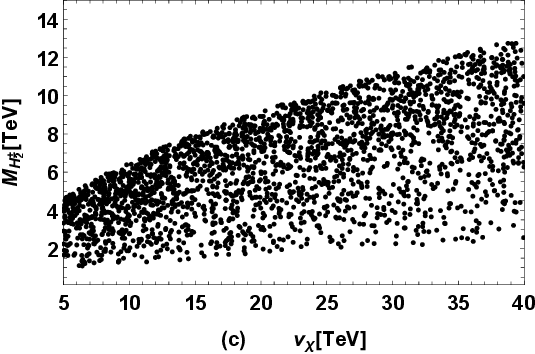}
\vspace{-0.2cm}
\caption[]{The results of charged Higgs masses $M_{H^\pm_1}$ (a), $M_{H^\pm_2}$ (b), $M_{H^\pm_2}$ (c) versus $\lambda_{10}$, $\kappa$, $v_\chi$ are plotted respectively by scanning the parameter space in Eq.~(\ref{eq25}).}
\label{MCHmass}
\end{figure}
Similar to the case of CP-odd Higgs, there are also two physical charged Higgs in the FDM. The results of charged Higgs masses $M_{H^\pm_1}$ versus $\lambda_{10}$, $M_{H^\pm_2}$ versus $\kappa$, $M_{H^\pm_2}$ versus $v_\chi$ are plotted in Fig.~\ref{MCHmass} (a), Fig.~\ref{MCHmass} (b), Fig.~\ref{MCHmass} (c) respectively. Fig.~\ref{MCHmass} (a) indicates that $M_{H^\pm_1}$ is mainly affected by $\lambda_{10}$, but the other scanning parameter can also influence the predicted $M_{H^\pm_1}$ especially for small $|\lambda_{10}|$. Similar to the case of $M_{A_2}$, Fig.~\ref{MCHmass} (b) and Fig.~\ref{MCHmass} (c) show that $M_{H^\pm_2}$ is also dominated by $\kappa$, $v_{\chi}$.

\section{The flavor changed neutral currents in the FDM}\label{sec4}

As shown in Eq.~(\ref{eq9}), the different generations of fermions couple to different Higgs bosons while $\Phi_3$ corresponding to the SM-like Higgs, which are quite different from the ones in the SM. In addition, the new defined $Z$ and $Z'$ gauge bosons can also mediate the FCNCs. Hence, observing the FCNCs in the FDM may be effective to test the model. In this section, we focus on the $B$ meson rare decay processes $\bar B \to X_s\gamma$, $B_s^0 \to \mu^+\mu^-$, the top quark rare decay processes $t\to ch$, $t\to uh$ and the charged lepton flavor violation processes $\tau\to 3e$, $\tau\to 3\mu$, $\mu\to 3e$ predicted in the FDM. And for simplicity, we take the nonzero $U_F(1)$ charge $z=1$ in the following analysis.

\subsection{$B$ meson rare decay processes $\bar B \to X_s\gamma$ and $B_s^0 \to \mu^+\mu^-$ in the FDM}\label{sec4-1}

The $B$ meson rare decay processes $\bar B \to X_s\gamma$, $B_s^0 \to \mu^+\mu^-$ are related closely to the NP contributions, and the average experimental data on the branching ratios of $\bar B \to X_s\gamma$, $B_s^0 \to \mu^+\mu^-$ are~\cite{ParticleDataGroup:2022pth}
\begin{eqnarray}
&&{\rm Br}(\bar B \to X_s\gamma)=(3.49\pm0.19)\times 10^{-4},\nonumber\\
&&{\rm Br}(B_s^0 \to \mu^+\mu^-)=(3.01\pm0.35)\times 10^{-9}.\label{eqBD}
\end{eqnarray}
The newly introduced scalars in the FDM including CP-even Higgs, CP-odd Higgs and charged Higgs can make contributions to these two processes, the analytical calculations of the contributions are collected in the appendix~\cite{Yang:2018fvw}.

\begin{figure}
\setlength{\unitlength}{1mm}
\centering
\includegraphics[width=2.7in]{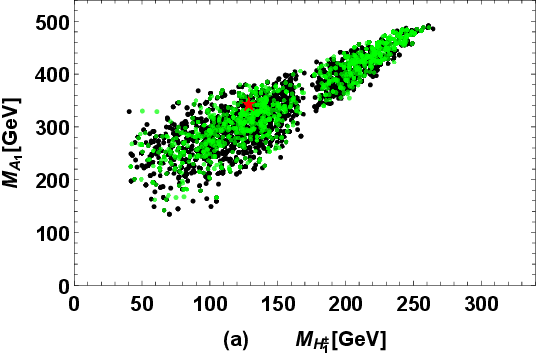}
\vspace{0.5cm}
\includegraphics[width=2.7in]{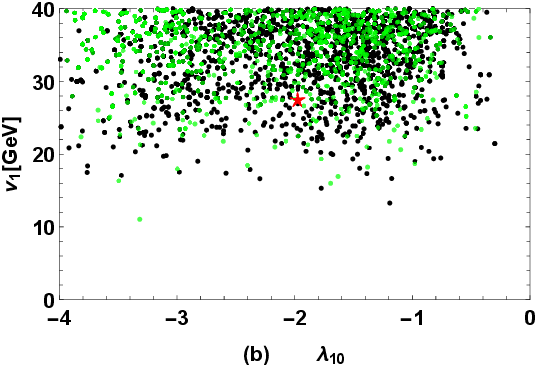}
\vspace{0cm}
\caption[]{Scanning the parameter spaces in Eq.~(\ref{eqSq}), Eq.~(\ref{eq25}), Eq.~(\ref{eq13}) and keeping $0.2243<|V_{us}|<0.2263$, $0.003516<|V_{ub}|<0.003716$, $0.4139<|V_{cb}|<0.4159$, $M_{H_2}$ in the range $124\;{\rm GeV}<M_{H_2}<126\;{\rm GeV}$, the allowed ranges of $M_{H^\pm_1}-M_{A_1}$ (a) and $v_1-\lambda_{10}$ (b) are plotted, where the black points, green points denote the results for ${\rm Br}(\bar B \to X_s\gamma)$ and ${\rm Br}(B_s^0 \to \mu^+\mu^-)$ in the experimental $2\sigma$, $1\sigma$ intervals respectively, the `red star' denotes the best fit with the B meson rare decay branching ratios in Eq.~(\ref{eqBD}) corresponding to $\chi^2=\mathbf{0.022}$.}
\label{Bdecay}
\end{figure}

Scanning the parameter spaces in Eq.~(\ref{eqSq}), Eq.~(\ref{eq25}) and the parameters $|m_{e,13}|$, $\theta_{e,ij},\;(ij=12,\;13,\;23)$ in the following range
\begin{eqnarray}
&&|m_{e,13}|=(0.0\sim0.3)\;{\rm GeV},\;\theta_{e,ij}=(-\pi\sim\pi),\label{eq13}
\end{eqnarray}
keeping $0.2243<|V_{us}|<0.2263$, $0.003516<|V_{ub}|<0.003716$, $0.4139<|V_{cb}|<0.4159$ and $M_{H_2}$ in the range $124\;{\rm GeV}<M_{H_2}<126\;{\rm GeV}$, we plot $M_{H^\pm_1}-M_{A_1}$, $v_1-\lambda_{10}$ in Fig.~\ref{Bdecay} (a), Fig.~\ref{Bdecay} (b) respectively, where the black points, green points denote the results for ${\rm Br}(\bar B \to X_s\gamma)$ and ${\rm Br}(B_s^0 \to \mu^+\mu^-)$ in the experimental $2\sigma$, $1\sigma$ intervals respectively, the `red star' denotes the best fit with the B meson rare decay branching ratios in Eq.~(\ref{eqBD}) corresponding to $\chi^2=0.022$. The results of this best fit are listed in Tab.~\ref{tab4}.
\begin{table*}
\begin{tabular*}{\textwidth}{@{\extracolsep{\fill}}llll@{}}
\hline
Observables & $O_i^{\rm th}$ & $O_i^{\rm exp}$ & Deviations in $\%$\\
\hline
$m_u$[MeV]                       & 2.154               & 2.16     & 0.28\\
$m_c$[GeV]                       & 1.658               & 1.67     & 0.72\\
$m_t$[GeV]                       & 172.5               & 172.5    & 0\\
$m_d$[MeV]                       & 4.67                & 4.67     & 0 \\
$m_s$[MeV]                       & 93.4                & 93.4     & 0 \\
$m_b$[GeV]                       & 4.78                & 4.78     & 0 \\
$|v_{us}|$                       & 0.2252              & 0.2253   & 0.044 \\
$|v_{ub}|$                       & 0.003617            & 0.003616 & 0.028 \\
$|v_{cb}|$                       & 0.4138              & 0.04149  & 0\\
${\rm Br}(\bar B \to X_s\gamma)$ & $3.49\times10^{-4}$ & $3.49\times10^{-4}$  & 0\\
${\rm Br}(B_s^0 \to \mu^+\mu^-)$ & $3.03\times10^{-9}$ & $3.01\times10^{-9}$  & 0.66\\
\hline
\end{tabular*}
\caption{Fit with the B meson rare decays: the results obtained for the best fit corresponding to $\chi^2=\mathbf{0.022}$.}
\label{tab4}
\end{table*}

For the parameters not shown in Fig.~\ref{Bdecay} such as $\lambda_1,\;\lambda_2,...$, they affect the predicted ${\rm Br}(\bar B \to X_s\gamma)$ and ${\rm Br}(B_s^0 \to \mu^+\mu^-)$ mildly. The results presented in Fig.~\ref{Bdecay} (a) indicate that $M_{H^\pm_1}$ is correlated strongly to $M_{A_1}$ because both of them mainly depend on $\lambda_{10}$ as shown in Fig.~\ref{MAmass} (a) and Fig.~\ref{MCHmass} (a). Eq.~(\ref{eqmu}) and Eq.~(\ref{eqmd}) show that the Yukawa couplings increase with decreasing $v_1$, i.e. the scalars in the FDM can make significant contributions to the $B$ meson rare decay processes $\bar B \to X_s\gamma$, $B_s^0 \to \mu^+\mu^-$ for small $v_1$, which leads to the experimental observations of ${\rm Br}(\bar B \to X_s\gamma)$ and ${\rm Br}(B_s^0 \to \mu^+\mu^-)$ prefer large $v_1$, and Fig.~\ref{Bdecay} (b) shows that $v_1$ is limited in the range $v_1\gtrsim15\;{\rm GeV}$.

\subsection{top quark rare decay processes $t\to ch$ and $t\to uh$}\label{sec4-2}

The branching ratios of the top quark rare decay processes $t\to ch$ and $t\to uh$ can be written as~\cite{YANG2018}
\begin{eqnarray}
&&{\rm Br}(t\rightarrow q_u h)=\frac{|\mathcal{M}_{t q_u h}|^2\sqrt{((m_t+m_h)^2-m_{q_u}^2)((m_t-m_h)^2-m_{q_u}^2)}}{32\pi m_t^3\Gamma^t_{{\rm total}}},
\end{eqnarray}
where $q_u=u,\;c$, the amplitude $\mathcal{M}_{tq_uh}$ can be read directly from the Yukawa couplings in Eq.~(\ref{eq9}), and $\Gamma^t_{{\rm total}}=1.42\;$GeV~\cite{ParticleDataGroup:2022pth} is the total decay width of top quark. The measured quark masses, CKM matrix and $B$ meson rare decay processes $\bar B \to X_s\gamma$, $B_s^0 \to \mu^+\mu^-$ should be considered in the calculations of top quark rare decay processes $t\to ch$ and $t\to uh$, hence we take the points obtained in Fig.~\ref{Bdecay} as inputs.

\begin{figure}
\setlength{\unitlength}{1mm}
\centering
\includegraphics[width=2.1in]{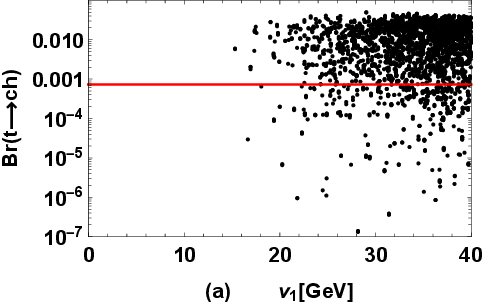}
\vspace{0.2cm}
\includegraphics[width=2.1in]{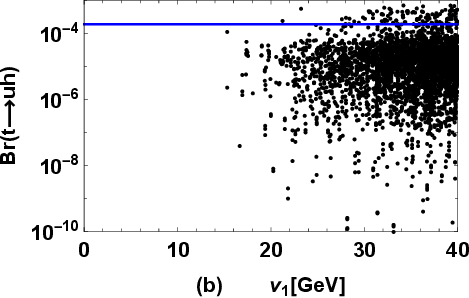}
\vspace{0.2cm}
\includegraphics[width=2.1in]{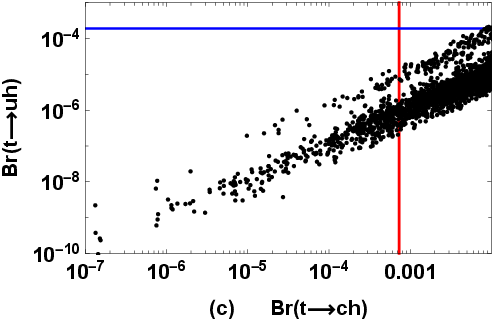}
\vspace{0cm}
\caption[]{Taking the points obtained in Fig.~\ref{Bdecay} as inputs, the results of ${\rm Br}(t\to ch)$ versus $|m_{u,13}|$ (a), $\theta_{u,12}$ (b) and the results of ${\rm Br}(t\to uh)$ versus $|m_{u,13}|$ (c), $\theta_{u,12}$ (d) are plotted, where red and blue lines denote the upper bounds on ${\rm Br}(t\to ch)$ and ${\rm Br}(t\to uh)$ from Particle Data Group~\cite{ParticleDataGroup:2022pth} respectively.}
\label{Fig2}
\end{figure}

Then we plot the results of ${\rm Br}(t\to ch)$ versus $v_1$ in Fig.~\ref{Fig2} (a), ${\rm Br}(t\to uh)$ versus $v_1$ in Fig.~\ref{Fig2} (b), and ${\rm Br}(t\to ch)$ versus ${\rm Br}(t\to uh)$ in Fig.~\ref{Fig2} (c). The red and blue lines in Fig.~\ref{Fig2} denote the upper bounds on ${\rm Br}(t\to ch)$ and ${\rm Br}(t\to uh)$ from Particle Data Group~\cite{ParticleDataGroup:2022pth} respectively. The picture illustrates that the results of ${\rm Br}(t\to ch)$, ${\rm Br}(t\to uh)$ obtained in the FDM can be large, which indicates that the processes $t\to ch$, $t\to uh$ have great opportunities to be observed experimentally. In addition, the parameter space of the model suffers constraints from the experimental upper bounds on ${\rm Br}(t\to ch)$ and ${\rm Br}(t\to uh)$.

\subsection{Lepton flavor violation processes $\tau\to 3e$, $\tau\to 3\mu$ and $\mu\to 3e$}\label{sec4-3}

Finally, we focus on the lepton flavor violation processes $\tau\to 3e$, $\tau\to 3\mu$, $\mu\to 3e$ predicted in the FDM. The corresponding amplitude can be written as~\cite{Hisano:1995cp}
\begin{eqnarray}
&&\mathcal{M}(e_j\rightarrow e_i e_i\bar e_i)=C_1^L\bar u_{e_i}(p_2)\gamma_\mu P_L u_{e_j}(p_1) u_{e_i}(p_3)\gamma^\mu P_L \nu_{e_i}(p_4)\nonumber\\
&&\qquad\quad+C_1^R\bar u_{e_i}(p_2)\gamma_\mu P_R u_{e_j}(p_1) u_{e_i}(p_3)\gamma^\mu P_R \nu_{e_i}(p_4)\nonumber\\
&&\qquad\quad+[C_2^L\bar u_{e_i}(p_2)\gamma_\mu P_L u_{e_j}(p_1) u_{e_i}(p_3)\gamma^\mu P_R \nu_{e_i}(p_4)\nonumber\\
&&\qquad\quad+C_2^R\bar u_{e_i}(p_2)\gamma_\mu P_R u_{e_j}(p_1) u_{e_i}(p_3)\gamma^\mu P_L \nu_{e_i}(p_4)-(p_2\leftrightarrow p_3)]\nonumber\\
&&\qquad\quad+[C_3^L\bar u_{e_i}(p_2) P_L u_{e_j}(p_1) u_{e_i}(p_3) P_L \nu_{e_i}(p_4)\nonumber\\
&&\qquad\quad+C_3^R\bar u_{e_i}(p_2) P_R u_{e_j}(p_1) u_{e_i}(p_3) P_R \nu_{e_i}(p_4)-(p_2\leftrightarrow p_3)],
\end{eqnarray}
where $i=1,\;2$ for $j=3$, $i=1$ for $j=2$, $u_{e_i}$ denotes the spinor of lepton, $\nu_{e_i}$ denotes the spinor of antilepton, $P_L=(1-\gamma_5)/2$, $P_R=(1+\gamma_5)/2$, and $p_k$ denotes the momentum of charged lepton with $k=1,2,3,4$. The coefficients $C_{1,2,3}^{L,R}$ from the contributions of Higgs bosons and $Z,\;Z'$ gauge bosons, can be obtained through the Yukawa couplings in Eq.~(\ref{eq9}) and the definition of covariant derivative in Eq.~(\ref{eqCD}). Then we can calculate the decay rate~\cite{Hisano:1995cp}
\begin{eqnarray}
&&\Gamma(e_j\rightarrow e_i e_i\bar e_i)=\frac{m_{e_j}^5}{1536\pi^3}\Big[\frac{1}{2}(|C_1^L|^2+|C_1^R|^2)+|C_2^L|^2+|C_2^R|^2+\frac{1}{8}(|C_3^L|^2+|C_3^R|^2)\Big].
\end{eqnarray}
The total decay widthes of $\mu,\;\tau$ are taken as $\Gamma^\mu_{{\rm total}}=2.996\times 10^{-19}\;$GeV, $\Gamma^\tau_{{\rm total}}=2.265\times 10^{-12}\;$GeV~\cite{ParticleDataGroup:2022pth}.

\begin{figure}
\setlength{\unitlength}{1mm}
\centering
\includegraphics[width=2.1in]{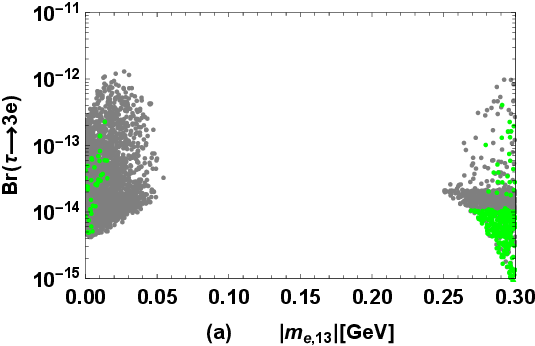}
\vspace{0.2cm}
\includegraphics[width=2.1in]{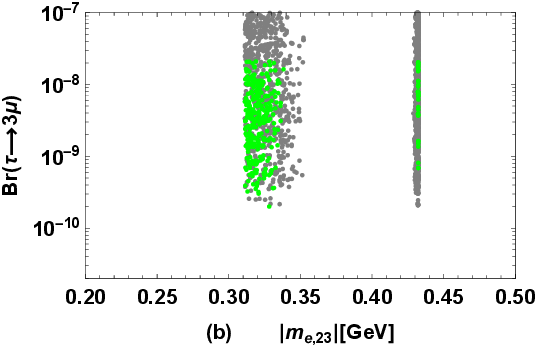}
\vspace{0.2cm}
\includegraphics[width=2.1in]{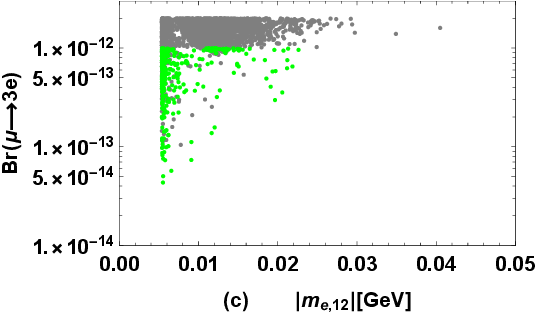}
\vspace{0.2cm}
\includegraphics[width=2.1in]{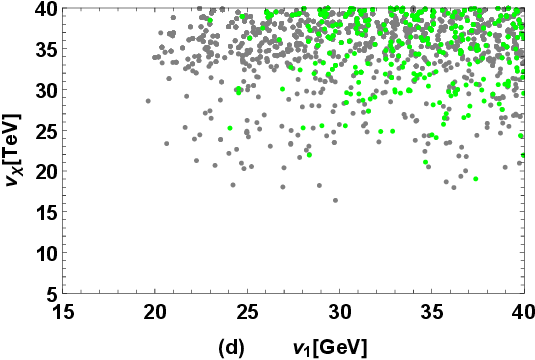}
\vspace{0.2cm}
\includegraphics[width=2.1in]{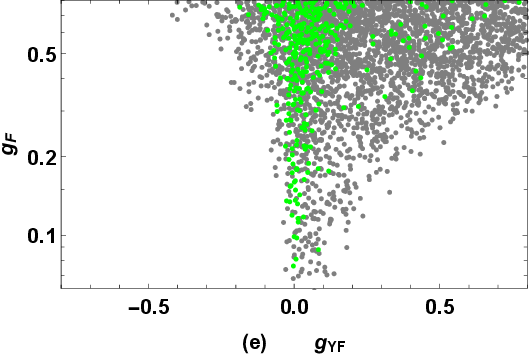}
\vspace{0.2cm}
\includegraphics[width=2.1in]{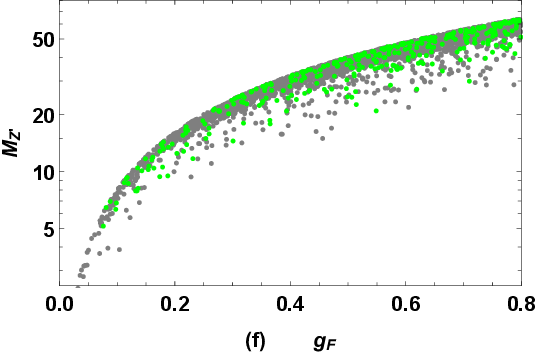}
\vspace{0cm}
\caption[]{Scanning the parameter space Eq.~(\ref{eq25}), Eq.~(\ref{eq13}), Eq.~(\ref{eq38}) and keeping $M_{H_2}$ in the range $124\;{\rm GeV}<M_{H_2}<126\;{\rm GeV}$ and ${\rm Br}(\tau\to 3e)<10^{-7}$, ${\rm Br}(\tau\to 3\mu)<10^{-7}$, ${\rm Br}(\mu\to 3e)<2\times10^{-12}$, the results of ${\rm Br}(\tau\to 3e)$ versus $|m_{e,13}|$ (a), ${\rm Br}(\tau\to 3\mu)$ versus $\theta_{e,23}$ (b), ${\rm Br}(\mu\to 3e)$ versus $\theta_{e,12}$ (c) and the allowed ranges of $v_\chi-v_1$ (d), $g_F-g_{YF}$ (e), $M_{Z'}-g_F$ (f) are plotted, where the gray points are excluded by the present limits, the green points denote the results which can reach future experimental sensitivities.}
\label{Fig3}
\end{figure}

Scanning the free parameter space in Eq.~(\ref{eq25}), Eq.~(\ref{eq13}) and
\begin{eqnarray}
&&g_{F}=(0,\;0.8),\;g_{F}=(-0.8,\;0.8),\label{eq38}
\end{eqnarray}
we plot the results of ${\rm Br}(\tau\to 3e)$ versus $|m_{e,13}|$, ${\rm Br}(\tau\to 3\mu)$ versus $\theta_{e,23}$, ${\rm Br}(\mu\to 3e)$ versus $\theta_{e,12}$ in Fig.~\ref{Fig3} (a), (b), (c) respectively by keeping $M_{H_2}$ in the range $124\;{\rm GeV}<M_{H_2}<126\;{\rm GeV}$ and ${\rm Br}(\tau\to 3e)<10^{-7}$, ${\rm Br}(\tau\to 3\mu)<10^{-7}$, ${\rm Br}(\mu\to 3e)<2\times10^{-12}$, then the allowed ranges of $v_\chi-v_1$, $g_F-g_{YF}$, $M_{Z'}-g_F$ are plotted in Fig.~\ref{Fig3} (d), (e), (f) respectively. The gray points in Fig.~\ref{Fig3} are excluded by the present limits, and the green points denote the results which can reach future experimental sensitivities, where the present limits and future sensitivities for the branching ratios of these LFV processes are listed in Tab.~\ref{tab5}.

\begin{table*}
\begin{tabular*}{\textwidth}{@{\extracolsep{\fill}}lll@{}}
\hline
Branching ratios                  & Present limit & Future sensitivity\\
\hline
${\rm Br}(\tau\to 3e)$            & $<2.7\times10^{-8}$    & $\sim10^{-10}$\\
${\rm Br}(\tau\to 3\mu)$          & $<2.1\times10^{-8}$    & $\sim10^{-10}$\\
${\rm Br}(\mu\to 3e)$             & $<10^{-12}$            & $\sim10^{-16}$\\
\hline
\end{tabular*}
\caption{Present limits and future sensitivities for the branching ratios for the LFV processes $\tau\to 3e$, $\tau\to 3\mu$ and $\mu\to 3e$~\cite{ParticleDataGroup:2022pth,Blondel:2013ia,Hayasaka:2013dsa}.}
\label{tab5}
\end{table*}

Fig.~\ref{Fig3} (a), (b), (c) show that the experimental upper bounds on ${\rm Br}(\tau\to 3\mu)$, ${\rm Br}(\mu\to 3e)$ limit the parameter space strictly while the predicted ${\rm Br}(\tau\to 3e)$ is less than about $10^{-12}$ which is hard to be observed in near future. In addition, the model predicts ${\rm Br}(\tau\to 3\mu)\gtrsim2\times 10^{-10}$, ${\rm Br}(\mu\to 3e)\gtrsim 5\times 10^{-14}$, which indicates observing the LFV processes $\tau\to 3\mu$ and $\mu\to 3e$ is also effective to test the FDM. It is obvious in Fig.~\ref{Fig3} (d) that the experimental upper bounds on the branching ratios of these LFV processes limit $v_1\gtrsim23\;{\rm GeV}$, $v_\chi\gtrsim20\;{\rm TeV}$. Fig.~\ref{Fig3} (e) indicates experimental constraints prefer $-0.2\lesssim g_{YF} \lesssim 0.2$ and $g_{YF} \lesssim -0.2$ is excluded completely. From Fig.~\ref{Fig3} (f), it can be seen explicitly that the allowed range of $M_{Z'}$ is related closely with the chosen value of $g_F$ and $M_{Z'}\gtrsim 5\;{\rm TeV}$.

\section{Summary}\label{sec5}

Motivated by the hierarchical structure of fermionic masses puzzle and fermionic flavor mixings puzzle, we propose a flavor-dependent model (FDM) to relate these two puzzles, i.e. the proposed FDM can explain the flavor mixings puzzle and mass hierarchy puzzle simultaneously. The model extends the SM by an extra $U(1)_F$ local gauge group, two scalar doublets, one scalar singlet and two right-handed neutrinos, where the new $U(1)_F$ charges are related to the particles' flavor. In the FDM, only the third generation of quarks and charged leptons achieve the masses at the tree level, the first two generations achieve masses through the mixings with the third generation, and the neutrinos obtain tiny Majorana masses through the so-called Type I see-saw mechanism. In addition, the $B$ meson rare decay processes $\bar B \to X_s\gamma$, $B_s^0 \to \mu^+\mu^-$, the top quark rare decay processes $t\to ch$, $t\to uh$ and the $\tau$ LFV processes $\tau\to 3e$, $\tau\to 3\mu$, $\mu\to 3e$ predicted in the FDM are analyzed. It is found that observing the top quark rare decay processes $t\to ch$, $t\to uh$ and the $\tau$ LFV decays $\tau\to 3\mu$, $\mu\to 3e$ is effective to test the FDM, while $\tau$ LFV decay $\tau\to 3e$ is hard to be observed experimentally. In addition, the model can fit the observed quark masses, CKM matrix, ${\rm Br}(\bar B \to X_s\gamma)$, ${\rm Br}(B_s^0 \to \mu^+\mu^-)$ well, and the VEVs of the two extra scalar doublets are limited to be larger than about $23\;{\rm GeV}$, new $Z'$ gauge boson is heavier than about $5\;{\rm TeV}$ and gauge kinetic mixing constant $g_{YF}$ is lager than $-0.2$ by considering the experimental upper bounds on the branching ratios of LFV decays $\tau\to 3\mu$, $\mu\to 3e$.

\appendix

\section{Contributions to $\bar B \to X_s\gamma$ and $B_s^0 \to \mu^+\mu^-$ in the FDM.\label{A1}}

Generally, the effective Hamilton for the transition $b\rightarrow s$ at hadronic scale can be written as
\begin{eqnarray}
&&H_{eff}=-\frac{4G_F}{\sqrt{2}}V_{ts}^\ast V_{tb}\Big[C_1\mathcal{O}^c_1+C_2\mathcal{O}_2^c+\sum_{i=3}^6\mathcal{O}_i+\sum_{i=7}^{10}(C_i\mathcal{O}_i+C'_i\mathcal{O}'_i)\nonumber\\
&&\qquad\;\quad\;+\sum_{i=S,P}(C_i\mathcal{O}_i+C'_i\mathcal{O}'_i)\Big],
\end{eqnarray}
where~\cite{O1,O2,Altmannshofer:2008dz,O3,O4,O6}
\begin{eqnarray}
&&{\cal O}_{_1}^u=(\bar{s}_{_L}\gamma_\mu T^au_{_L})(\bar{u}_{_L}\gamma^\mu T^ab_{_L})\;,\;\;
{\cal O}_{_2}^u=(\bar{s}_{_L}\gamma_\mu u_{_L})(\bar{u}_{_L}\gamma^\mu b_{_L})\;,
\nonumber\\
&&{\cal O}_{_3}=(\bar{s}_{_L}\gamma_\mu b_{_L})\sum\limits_q(\bar{q}\gamma^\mu q)\;,\;\;
{\cal O}_{_4}=(\bar{s}_{_L}\gamma_\mu T^ab_{_L})\sum\limits_q(\bar{q}\gamma^\mu T^aq)\;,
\nonumber\\
&&{\cal O}_{_5}=(\bar{s}_{_L}\gamma_\mu\gamma_\nu\gamma_\rho b_{_L})\sum\limits_q(\bar{q}\gamma^\mu
\gamma^\nu\gamma^\rho q)\;,\;\;
{\cal O}_{_6}=(\bar{s}_{_L}\gamma_\mu\gamma_\nu\gamma_\rho T^ab_{_L})\sum\limits_q(\bar{q}\gamma^\mu
\gamma^\nu\gamma^\rho T^aq)\;,
\nonumber\\
&&{\cal O}_{_7}={e\over 16\pi^2}m_{_b}(\bar{s}_{_L}\sigma_{_{\mu\nu}}b_{_R})F^{\mu\nu}\;,\;\;
{\cal O}_{_7}'={e\over 16\pi^2}m_{_b}(\bar{s}_{_R}\sigma_{_{\mu\nu}}b_{_L})F^{\mu\nu}\;,\;\;
\nonumber\\
&&{\cal O}_{_8}={g_{_s}\over 16\pi^2}m_{_b}(\bar{s}_{_L}\sigma_{_{\mu\nu}}T^ab_{_R})G^{a,\mu\nu}\;,\;\;
{\cal O}_{_8}'={g_{_s}\over 16\pi^2}m_{_b}(\bar{s}_{_R}\sigma_{_{\mu\nu}}T^ab_{_L})G^{a,\mu\nu}\;,\;\;
\nonumber\\
&&{\cal O}_{_9}={e^2\over g_{_s}^2}(\bar{s}_{_L}\gamma_\mu b_{_L})\bar{l}\gamma^\mu l\;,\;\;
{\cal O}_{_9}'={e^2\over g_{_s}^2}(\bar{s}_{_R}\gamma_\mu b_{_R})\bar{l}\gamma^\mu l\;,\;\;
\nonumber\\
&&{\cal O}_{_{10}}={e^2\over g_{_s}^2}(\bar{s}_{_L}\gamma_\mu b_{_L})\bar{l}\gamma^\mu\gamma_5 l\;,\;\;
{\cal O}_{_{10}}'={e^2\over g_{_s}^2}(\bar{s}_{_R}\gamma_\mu b_{_R})\bar{l}\gamma^\mu\gamma_5 l\;,\;\;
\nonumber\\
&&{\cal O}_{_S}={e^2\over16\pi^2}m_{_b}(\bar{s}_{_L}b_{_R})\bar{l}l\;,\;\;
{\cal O}_{_S}'={e^2\over16\pi^2}m_{_b}(\bar{s}_{_R}b_{_L})\bar{l}l\;,\;\;\nonumber\\
&&{\cal O}_{_P}={e^2\over16\pi^2}m_{_b}(\bar{s}_{_L}b_{_R})\bar{l}\gamma_5l\;,\;\;
{\cal O}_{_P}'={e^2\over16\pi^2}m_{_b}(\bar{s}_{_R}b_{_L})\bar{l}\gamma_5l.
\label{operators}
\end{eqnarray}
In the definitions above, $g_s$ is the strong coupling constant, $T^a\,(a=1,...,8)$ are $SU(3)$ generators, $F^{\mu\nu}$ and $G^{\mu\nu}$ are the electromagnetic and gluon field strength tensors respectively.

\begin{figure}
\setlength{\unitlength}{1mm}
\centering
\includegraphics[width=4in]{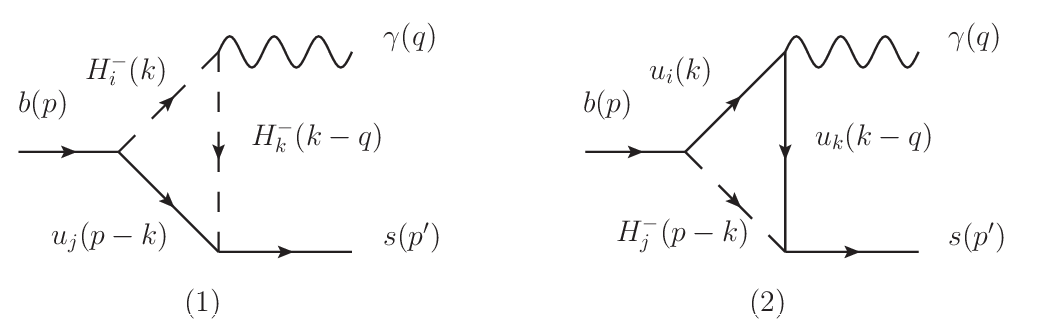}
\vspace{0cm}
\caption[]{The  one loop Feynman diagrams contributing to $\bar{B}\rightarrow X_s\gamma$ from charged Higgs in the FDM.}
\label{figBsr}
\end{figure}
The dominant contributions to $b\rightarrow s\gamma$ come from the charged Higgs in the FDM, and the leading-order Feynman diagrams are plotted in Fig.~\ref{figBsr}. Then the branching ratio of $\bar{B}\rightarrow X_s\gamma$ can be written as
\begin{eqnarray}
&&Br(\bar{B}\rightarrow X_s\gamma)=R\Big(|C_{7\gamma}(\mu_b)|^2+N(E_\gamma)\Big)\;,
\end{eqnarray}
where the overall factor $R=2.47\times10^{-3}$, and the nonperturbative contribution $N(E_\gamma)=(3.6\pm0.6)\times10^{-3}$\cite{H5}. $C_{7\gamma}(\mu_b)$ can be written as
\begin{eqnarray}
&&C_{7\gamma}(\mu_b)=C_{7\gamma,SM}(\mu_b)+C_{7,NP}(\mu_b),
\end{eqnarray}
where the hadron scale $\mu_b=2.5$ GeV and $C_{7\gamma,SM}(\mu_b) = -0.3689$ for the SM contribution at NNLO level~\cite{H5,H6,H7,H8}. In new physics models, the corresponding Wilson coefficients at the bottom quark scale are~\cite{H9,H10}
\begin{eqnarray}
&&C_{7,NP}(\mu_b)\approx0.5696
C_{7,NP}(\mu_{EW})+0.1107 C_{8,NP}(\mu_{EW}),
\end{eqnarray}
where
\begin{eqnarray}
&&C_{7,NP}^{NP}(\mu_{EW})=C_{7,NP}^{(1)}(\mu_{EW})+C_{7,NP}^{(2)}(\mu_{EW})+C_{7,NP}^{\prime(1)}(\mu_{EW})+C_{7,NP}^{\prime(2)}(\mu_{EW}),\nonumber\\
&&C_{8,NP}(\mu_{EW})=C_{8g,NP}(\mu_{EW})+C_{8g,NP}^{\prime}(\mu_{EW}),
\end{eqnarray}
The coefficients $C_{7,NP}^{(1,2)}(\mu_{EW})$ are Wilson coefficients of the process $b\rightarrow s\gamma$ and can be calculated from the diagrams in Fig.~\ref{figBsr} (1), (2) respectively, the results read
\begin{eqnarray}
&&C_{7,NP}^{(1)}(\mu_{EW})=\sum_{H^-_i,u_j}\frac{s_W^2}{2e^2V^*_{ts}V_{tb}} \Big\{ \frac{1}{2}C_{H^-_i\bar s u_j}^R C_{H^-_i b \bar u_j}^{L}[-I_3(x_{u_j},x_{H^-_i})+I_4(x_{u_j},x_{H^-_i})]+\nonumber\\
&&\qquad\qquad\qquad\quad\frac{m_{u_j}}{m_b}C_{H^-_i\bar s u_j}^L C_{H^-_i b \bar u_j}^{L}[-I_1(x_{u_j},x_{H^-_i})+I_3(x_{u_j},x_{H^-_i})]\Big\},\nonumber\\
&&C_{7,NP}^{(2)}(\mu_{EW})=\sum_{H^-_j,u_i}\frac{s_W^2}{3e^2V^*_{ts}V_{tb}} \Big\{ \frac{1}{2}C_{H^-_j\bar s u_i}^R C_{H^-_j b \bar u_i}^{L}[-I_1(x_{u_i},x_{H^-_j})+2I_3(x_{u_i},x_{H^-_j})\nonumber\\
&&\qquad\qquad\qquad\quad-I_4(x_{u_i},x_{H^-_j})]+\frac{m_{u_i}}{m_b}C_{H^-_j\bar s u_i}^L C_{H^-_j b \bar u_i}^{L}[I_1(x_{u_i},x_{H^-_j})-I_2(x_{u_i},x_{H^-_j})\nonumber\\
&&\qquad\qquad\qquad\quad-I_3(x_{u_i},x_{H^-_j})]\Big\},\nonumber\\
&&C_{7}^{\prime NP(a)}(\mu_{EW})=C_{7}^{\prime NP(a)}(\mu_{EW})(L\leftrightarrow R), (a=1,2),
\end{eqnarray}
where $x_i=\frac{m_i^2}{m_W^2}$, $C_{abc}^{L,R}$ denotes the scalar parts of the interaction vertex about $abc$ with $a, b, c$ denoting the interactional particles, and the loop integral functions $I_{1,...,4}$ can be found in our previous work~\cite{Yang:2018fvw}. In addition, $C_{8g,NP}(\mu_{EW})$ and $C_{8g,NP}^{\prime}(\mu_{EW})$ at electroweak scale are
\begin{eqnarray}
&&C_{8g,NP}(\mu_{EW})=[C_{7,NP}^{(2)}(\mu_{EW})+C_{7,NP}^{(3)}(\mu_{EW})]/Q_u,\nonumber\\
&&C_{8g,NP}^{\prime}(\mu_{EW})=C_{8g,NP}(\mu_{EW})(L\leftrightarrow R),
\end{eqnarray}
where $Q_u=2/3$.

\begin{figure}
\setlength{\unitlength}{1mm}
\centering
\includegraphics[width=5.5in]{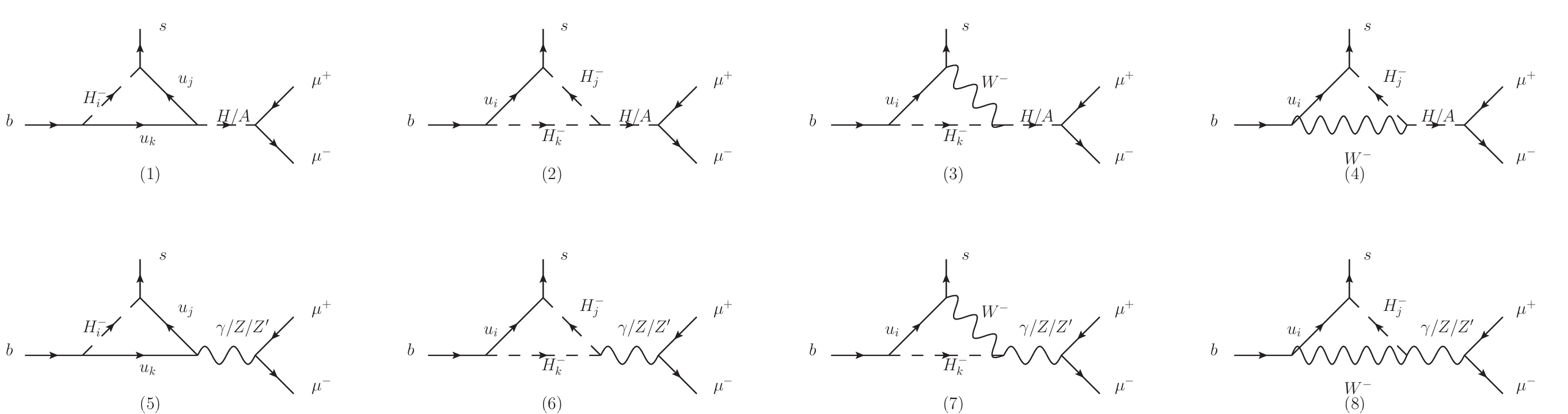}
\vspace{0cm}
\caption[]{The Feynman diagrams contributing to the decay $B_s^0\rightarrow\mu^+\mu^-$ in the B-LSSM} \label{Bmumu}
\end{figure}

The main Feynman diagrams contributing to $B_s^0 \to \mu^+\mu^-$ are plotted in Fig.~\ref{Bmumu}.
At the electroweak energy scale $\mu_{EW}$, the corresponding Wilson coefficients can be written as
\begin{eqnarray}
&&C_{_{S,NP}}(\mu_{_{\rm EW}})=\frac{\sqrt{2}s_{_W}c_{_W}}{4m_be^3V_{ts}^*V_{tb}}\Big[C_{_{S,NP}}^{(1)}(\mu_{_{\rm EW}})+C_{_{S,NP}}^{(2)}(\mu_{_{\rm EW}})+C_{_{S,NP}}^{(3)}(\mu_{_{\rm EW}})+C_{_{S,NP}}^{(4)}(\mu_{_{\rm EW}})\nonumber\\
&&\qquad\;\qquad\;\qquad+C_{_{S,NP}}^{(6)}(\mu_{_{\rm EW}})\Big],\nonumber\\
&&C_{_{S,NP}}^\prime(\mu_{_{\rm EW}})=C_{_{S,NP}}(\mu_{_{\rm EW}})(L\leftrightarrow R),\nonumber\\
&&C_{_{P,NP}}(\mu_{_{\rm EW}})=\frac{\sqrt{2}s_{_W}c_{_W}}{4m_be^3V_{ts}^*V_{tb}}\Big[C_{_{P,NP}}^{(1)}(\mu_{_{\rm EW}})+C_{_{P,NP}}^{(2)}(\mu_{_{\rm EW}})+C_{_{P,NP}}^{(3)}(\mu_{_{\rm EW}})+C_{_{P,NP}}^{(4)}(\mu_{_{\rm EW}})\nonumber\\
&&\qquad\;\qquad\;\qquad+C_{_{P,NP}}^{(6)}(\mu_{_{\rm EW}})\Big],\nonumber\\
&&C_{_{P,NP}}^\prime(\mu_{_{\rm EW}})=-C_{_{P,NP}}(\mu_{_{\rm EW}})(L\leftrightarrow R),\nonumber\\
&&C_{_{9,NP}}(\mu_{_{\rm EW}})=\frac{\sqrt{2}s_{_W}c_{_W}g_{_s}^2}{64\pi^2e^3V_{ts}^*V_{tb}}\Big[C_{_{9,NP}}^{(5)}(\mu_{_{\rm EW}})+C_{_{9,NP}}^{(6)}(\mu_{_{\rm EW}})+C_{_{9,NP}}^{(7)}(\mu_{_{\rm EW}})+C_{_{9,NP}}^{(8)}(\mu_{_{\rm EW}})\Big]\;,\nonumber\\
&&C_{_{9,NP}}^\prime(\mu_{_{\rm EW}})=C_{_{9,NP}}(\mu_{_{\rm EW}})(L\leftrightarrow R),\nonumber\\
&&C_{_{10,NP}}(\mu_{_{\rm EW}})=\frac{\sqrt{2}s_{_W}c_{_W}g_{_s}^2}{64\pi^2e^3V_{ts}^*V_{tb}}\Big[C_{_{10,NP}}^{(5)}(\mu_{_{\rm EW}})+C_{_{10,NP}}^{(6)}(\mu_{_{\rm EW}})+C_{_{10,NP}}^{(7)}(\mu_{_{\rm EW}})+C_{_{10,NP}}^{(8)}(\mu_{_{\rm EW}})\Big]\;,\nonumber\\
&&C_{_{10,NP}}^\prime(\mu_{_{\rm EW}})=-C_{_{10,NP}}(\mu_{_{\rm EW}})(L\leftrightarrow R).
\label{Wilson-Coefficients1}
\end{eqnarray}
The superscripts $(1,...,8)$ corresponding to the contributions in Fig.~\ref{Bmumu} (1,...,8) respectively and the results can be written as
\begin{eqnarray}
&&C_{_{S,NP}}^{(1)}(\mu_{_{\rm EW}})=\sum_{H^-_i,u_j,u_k}^{S=H_l,A_l}\frac{C_{\bar\mu S\mu }^L+C_{\bar\mu S\mu}^R}{2(m_b^2-m_S^2)}\Big[C_{H^-_i\bar s u_j}^R C_{\bar u_j S u_k}^L C_{\bar u_k b H^-_i}^R G_2(x_{\tilde H^\pm_i},x_{u_j},x_{u_k})\nonumber\\
&&\qquad\qquad\qquad+m_{u_j}m_{u_k}C_{H^-_i\bar s u_j}^R C_{\bar u_j S u_k}^R C_{\bar u_k b H^-_i}^R G_1(x_{\tilde H^\pm_i},x_{u_j},x_{u_k})\Big],\nonumber\\
&&C_{_{P,NP}}^{(1)}(\mu_{_{\rm EW}})=\sum_{H^-_i,u_j,u_k}^{S=H_l,A_l}\frac{-C_{\bar\mu S\mu}^L+C_{\bar\mu S\mu}^R}{2(m_b^2-m_S^2)}\Big[C_{H^-_i\bar s u_j}^R C_{\bar u_j S u_k}^L C_{\bar u_k b H^-_i}^R G_2(x_{\tilde H^\pm_i},x_{u_j},x_{u_k})\nonumber\\
&&\qquad\qquad\qquad+m_{u_j}m_{u_k}C_{H^-_i\bar s u_j}^R C_{\bar u_j S u_k}^R C_{\bar u_k b H^-_i}^R G_1(x_{\tilde H^\pm_i},x_{u_j},x_{u_k})\Big],
\end{eqnarray}
\begin{eqnarray}
&&C_{_{S,NP}}^{(2)}(\mu_{_{\rm EW}})=\sum_{u_i,H^\pm_j,H^\pm_k}^{S=H_l,A_l}\frac{1}{2(m_b^2-m_S^2)}m_{u_i}C_{\bar s u_i H^\pm_j}^RC_{\bar u_i b H^\pm_k}^RC_{S H^\pm_j H^\pm_k}G_1(x_{u_i},x_{H^\pm_j},x_{H^\pm_k})\nonumber\\
&&\qquad\qquad\qquad \times(C_{\bar\mu S\mu}^L+C_{\bar\mu S\mu}^R),\nonumber\\
&&C_{_{p,NP}}^{(2)}(\mu_{_{\rm EW}})=\sum_{u_i,H^\pm_j,H^\pm_k}^{S=H_l,A_l}\frac{1}{2(m_b^2-m_S^2)}m_{u_i}C_{\bar s u_i H^\pm_j}^RC_{\bar u_i b H^\pm_k}^RC_{S H^\pm_j H^\pm_k}G_1(x_{u_i},x_{H^\pm_j},x_{H^\pm_k})\nonumber\\
&&\qquad\qquad\qquad \times(-C_{\bar\mu S\mu}^L+C_{\bar\mu S\mu}^R),
\end{eqnarray}
\begin{eqnarray}
&&C_{_{S,NP}}^{(3)}(\mu_{_{\rm EW}})=\sum_{u_i,H^\pm_k}^{S=H_l,A_l}\frac{-C_{W^\pm S H^\pm_k}}{2(m_b^2-m_S^2)}\Big[C_{\bar s W^\pm u_i}^LC_{\bar u_i H^\pm_k b}^RG_2(x_{u_i},1,x_{H^\pm_k})-2m_bm_{u_i}C_{\bar s W^\pm u_i}^L\nonumber\\
&&\qquad\qquad\qquad \times C_{\bar u_i H^\pm_k b}^LG_1(x_{u_i},1,x_{H^\pm_k})\Big](C_{\bar\mu S\mu}^L+C_{\bar\mu S\mu}^R),\nonumber\\
&&C_{_{P,NP}}^{(3)}(\mu_{_{\rm EW}})=\sum_{u_i,H^\pm_k}^{S=H_l,A_l}\frac{-C_{W^\pm S H^\pm_k}}{2(m_b^2-m_S^2)}\Big[C_{\bar s W^\pm u_i}^LC_{\bar u_i H^\pm_k b}^RG_2(x_{u_i},1,x_{H^\pm_k})-2m_bm_{u_i}C_{\bar s W^\pm u_i}^L\nonumber\\
&&\qquad\qquad\qquad \times C_{\bar u_i H^\pm_k b}^LG_1(x_{u_i},1,x_{H^\pm_k})\Big](-C_{\bar\mu S\mu}^L+C_{\bar\mu S\mu}^R),
\end{eqnarray}
\begin{eqnarray}
&&C_{_{S,NP}}^{(4)}(\mu_{_{\rm EW}})=\sum_{u_i,H^\pm_j}^{S=H_l,A_l}\frac{-C_{W^\pm S H^\pm_j}}{2(m_b^2-m_S^2)}C_{\bar s H^\pm_j u_i}^RC_{\bar u_i W^\pm b}^RG_2(x_{u_i},x_{H^\pm_j},1)(C_{\bar\mu S\mu}^L+C_{\bar\mu S\mu}^R),\nonumber\\
&&C_{_{S,NP}}^{(4)}(\mu_{_{\rm EW}})=\sum_{u_i,H^\pm_j}^{S=H_l,A_l}\frac{-C_{W^\pm S H^\pm_j}}{2(m_b^2-m_S^2)}C_{\bar s H^\pm_j u_i}^RC_{\bar u_i W^\pm b}^RG_2(x_{u_i},x_{H^\pm_j},1)(-C_{\bar\mu S\mu}^L+C_{\bar\mu S\mu}^R),\nonumber\\
\end{eqnarray}
\begin{eqnarray}
&&C_{_{9,NP}}^{(5)}(\mu_{_{\rm EW}})=\sum_{\tilde H^\pm_i,u_j,u_k
}^{V}\frac{C_{\bar\mu V\mu}^L+C_{\bar\mu V\mu}^R}{-2(m_b^2-m_V^2)}\Big[-\frac{1}{2}C_{H^\pm_i\bar s u_j}^R C_{\bar u_j V u_k}^R C_{u_k s H^\pm_i}^L G_2(x_{H^\pm_i},x_{u_j},x_{u_k})\nonumber\\
&&\qquad\qquad\qquad+m_{u_j}m_{u_k} C_{H^\pm_i\bar s u_j}^R C_{\bar u_j V u_k}^L C_{\bar u_k s H^\pm_i}^L G_1(x_{H^\pm_i},x_{u_j},x_{u_k})\Big],\nonumber\\
&&C_{_{10,NP}}^{(5)}(\mu_{_{\rm EW}})=\sum_{\tilde H^\pm_i,u_j,u_k
}^{V}\frac{-C_{\bar\mu V\mu}^L+C_{\bar\mu V\mu}^R}{-2(m_b^2-m_V^2)}\Big[-\frac{1}{2}C_{H^\pm_i\bar s u_j}^R C_{\bar u_j V u_k}^R C_{u_k s H^\pm_i}^L G_2(x_{H^\pm_i},x_{u_j},x_{u_k})\nonumber\\
&&\qquad\qquad\qquad+m_{u_j}m_{u_k} C_{H^\pm_i\bar s u_j}^R C_{\bar u_j V u_k}^L C_{\bar u_k s H^\pm_i}^L G_1(x_{H^\pm_i},x_{u_j},x_{u_k})\Big],
\end{eqnarray}
\begin{eqnarray}
&&C_{_{9,NP}}^{(6)}(\mu_{_{\rm EW}})=\sum_{u_i,H^\pm_j,H^\pm_k}^V\frac{C_{\bar\mu V\mu}^L+C_{\bar\mu V\mu}^R}{4(m_b^2-m_V^2)}C_{\bar s u_i H^\pm_j}^RC_{\bar u_i b H^\pm_k}^LC_{V H^\pm_j H^\pm_k}G_2(x_{u_i},x_{H^\pm_j},x_{H^\pm_k}),\nonumber\\
&&C_{_{10,NP}}^{(6)}(\mu_{_{\rm EW}})=\sum_{u_i,H^\pm_j,H^\pm_k}^V\frac{-C_{\bar\mu V\mu}^L+C_{\bar\mu V\mu}^R}{4(m_b^2-m_V^2)}C_{\bar s u_i H^\pm_j}^RC_{\bar u_i b H^\pm_k}^LC_{V H^\pm_j H^\pm_k}G_2(x_{u_i},x_{H^\pm_j},x_{H^\pm_k}),\nonumber\\
&&C_{_{S,NP}}^{(6)}(\mu_{_{\rm EW}})=\sum_{u_i,H^\pm_j,H^\pm_k}^V\frac{C_{\bar\mu V\mu}^L+C_{\bar\mu V\mu}^R}{-2(m_b^2-m_V^2)}m_bm_{u_i}C_{\bar s u_i H^\pm_j}^RC_{\bar u_i b H^\pm_k}^RC_{V H^\pm_j H^\pm_k}G_1(x_{u_i},x_{H^\pm_j},x_{H^\pm_k}),\nonumber\\
&&C_{_{P,NP}}^{(6)}(\mu_{_{\rm EW}})=\sum_{u_i,H^\pm_j,H^\pm_k}^V\frac{C_{\bar\mu V\mu}^L-C_{\bar\mu V\mu}^R}{-2(m_b^2-m_V^2)}m_bm_{u_i}C_{\bar s u_i H^\pm_j}^RC_{\bar u_i b H^\pm_k}^RC_{V H^\pm_j H^\pm_k}G_1(x_{u_i},x_{H^\pm_j},x_{H^\pm_k}),\nonumber\\
\end{eqnarray}
\begin{eqnarray}
&&C_{_{9,NP}}^{(7)}(\mu_{_{\rm EW}})=\sum_{u_i,H^\pm_k}^V\frac{C_{\bar\mu V\mu}^L+C_{\bar\mu V\mu}^R}{2(m_b^2-m_V^2)}m_{u_i}C_{\bar s u_i W^\pm}^LC_{\bar u_i b H^\pm_k}^LC_{V W^\pm H^\pm_k}G_1(x_{u_i},x_W,x_{H^\pm_k}),\nonumber\\
&&C_{_{10,NP}}^{(7)}(\mu_{_{\rm EW}})=\sum_{u_i,H^\pm_k}^V\frac{-C_{\bar\mu V\mu}^L+C_{\bar\mu V\mu}^R}{2(m_b^2-m_V^2)}m_{u_i}C_{\bar s u_i W^\pm}^LC_{\bar u_i b H^\pm_k}^LC_{V W^\pm H^\pm_k}G_1(x_{u_i},x_W,x_{H^\pm_k}),\nonumber\\
\end{eqnarray}
\begin{eqnarray}
&&C_{_{9,NP}}^{(8)}(\mu_{_{\rm EW}})=\sum_{u_i,H^\pm_j}^V\frac{C_{\bar\mu V\mu}^L+C_{\bar\mu V\mu}^R}{2(m_b^2-m_V^2)}m_{u_i}C_{\bar s u_i H^\pm_j}^RC_{\bar u_i b W^\pm}^LC_{V W^\pm H^\pm_k}G_1(x_{u_i},x_{H^\pm_j},x_W),\nonumber\\
&&C_{_{10,NP}}^{(8)}(\mu_{_{\rm EW}})=\sum_{u_i,H^\pm_j}^V\frac{-C_{\bar\mu V\mu}^L+C_{\bar\mu V\mu}^R}{2(m_b^2-m_V^2)}m_{u_i}C_{\bar s u_i H^\pm_j}^RC_{\bar u_i b W^\pm}^LC_{V W^\pm H^\pm_k}G_1(x_{u_i},x_{H^\pm_j},x_W),\nonumber\\
\end{eqnarray}
where $V$ denotes the vector bosons $\gamma$, $Z$, $Z'$, and $C_{aVb}$ denote the scalar parts of the corresponding interaction vertex $aVb$. The concrete expressions for loop integral $G_k(k=1,...,4)$ can be found in the Appendix B of our previous work~\cite{Yang:2018fvw}. The Wilson coefficients at hadronic energy scale from the SM to next-to-next-to-logarithmic accuracy are shown in Table I.
\begin{table}
\begin{tabular}{|c|c|c|c|}
\hline
\hline
$C_{_7}^{eff,SM}$    & $C_{_8}^{eff,SM}$    & $C_{_9}^{eff,SM}$ & $C_{_{10}}^{eff,SM}$\\
\hline
$-0.304$   & $-0.167$  & $4.211$ & $-4.103$\\
\hline
\hline
\end{tabular}
\caption{At hadronic scale $\mu=m_{_b}$, SM Wilson coefficients to next-to-next-to-logarithmic accuracy. \label{tab1}}
\end{table}
In addition, the Wilson coefficients in Eq.~(\ref{Wilson-Coefficients1}) should be evolved down to hadronic scale $\mu\sim m_b$ by the renormalization group equations:
\begin{eqnarray}
&&\overrightarrow{C}_{_{NP}}(\mu)=\widehat{U}(\mu,\mu_0)\overrightarrow{C}_{_{NP}}(\mu_0)
\;,\nonumber\\
&&\overrightarrow{C^\prime}_{_{NP}}(\mu)=\widehat{U^\prime}(\mu,\mu_0)
\overrightarrow{C^\prime}_{_{NP}}(\mu_0)
\label{evaluation1}
\end{eqnarray}
with
\begin{eqnarray}
&&\overrightarrow{C}_{_{NP}}^{T}=\Big(C_{_{1,NP}},\;\cdots,\;C_{_{6,NP}},
C_{_{7,NP}}^{eff},\;C_{_{8,NP}}^{eff},\;C_{_{9,NP}}^{eff}-Y(q^2),\;
C_{_{10,NP}}^{eff}\Big)
\;,\nonumber\\
&&\overrightarrow{C}_{_{NP}}^{\prime,\;T}=\Big(C_{_{7,NP}}^{\prime,\;eff},\;
C_{_{8,NP}}^{\prime,\;eff},\;C_{_{9,NP}}^{\prime,\;eff},\;
C_{_{10,NP}}^{\prime,\;eff}\Big)\;.
\label{evaluation2}
\end{eqnarray}
Correspondingly, the evolving matrices $\widehat{U}(\mu,\mu_0),\;\widehat{U^\prime}(\mu,\mu_0)$ can be found in our previous work~\cite{Yang:2018fvw}.

Then, the squared amplitude can be written as
\begin{eqnarray}
&&|\mathcal{M}_s|^2=16G_F^2|V_{tb}V_{ts}^*|^2M_{B_s^0}^2\Big[|F_S^s|^2+|F_P^s+2m_{\mu}F_A^s|^2\Big],
\end{eqnarray}
and
\begin{eqnarray}
&&F_S^s=\frac{\alpha_{EW}(\mu_b)}{8\pi}\frac{m_b M_{B_s^0}^2}{m_b+m_s}f_{B_s^0}(C_S-C_S'),\\
&&F_P^s=\frac{\alpha_{EW}(\mu_b)}{8\pi}\frac{m_b M_{B_s^0}^2}{m_b+m_s}f_{B_s^0}(C_P-C_P'),\\
&&F_A^s=\frac{\alpha_{EW}(\mu_b)}{8\pi}f_{B_s^0}\Big[C_{10}^{eff}(\mu_b)-C_{10}^{\prime eff}(\mu_b)\Big],
\end{eqnarray}
where $f_{B_s^0}=(227\pm8)\;{\rm MeV}$ denote the decay constants, $M_{B_s^0}=5.367\; {\rm GeV}$ denote the masses of neutral meson $B_s^0$. The branching ratio of $B_s^0\rightarrow\mu^+\mu^-$ can be written as
\begin{eqnarray}
&&Br(B_s^0\rightarrow\mu^+\mu^-)=\frac{\tau_{B_s^0}}{16\pi}\frac{|\mathcal{M}_s|^2}{M_{B_s^0}}\sqrt{1-\frac{4m_{\mu}^2}{M_{B_s^0}^2}},
\end{eqnarray}
with $\tau_{B_s^0}=1.466(31)\;{\rm ps}$ denoting the life time of meson.

\begin{acknowledgments}

The work has been supported by the National Natural Science Foundation of China (NNSFC) with Grants No. 12075074, No. 12235008, Hebei Natural Science Foundation with Grant No. A2022201017, No. A2023201041, Natural Science Foundation of Guangxi Autonomous Region with Grant No. 2022GXNSFDA035068, the youth top-notch talent support program of the Hebei Province.

\end{acknowledgments}

\end{document}